\documentclass[floatfix,aps,preprint,prb,singlecolumn,superscriptaddress,longbibliography]{revtex4-2}
\usepackage{multirow}
\usepackage{caption}
\usepackage{ragged2e}
\usepackage{setspace}
\usepackage{graphicx}
\usepackage{siunitx}
\usepackage{color}
\usepackage{float}
\usepackage{amsmath}
\usepackage{makecell}
\usepackage[breaklinks]{hyperref}
\hypersetup{
	pdfnewwindow=true,      
    colorlinks=true,       
    linkcolor=blue,          
    citecolor=blue,        
    filecolor=blue,      
    urlcolor=blue           
}

\usepackage{mhchem} 
\usepackage{chemformula} 

\usepackage{amsmath, amsfonts, amssymb}

\begin{document}
\title{Substrate influence on transition metal dichalcogenide monolayer exciton absorption linewidth broadening}

\author{Fuhui Shao}
\affiliation{Universit\'e Paris-Saclay, CNRS, Laboratoire de Physique des Solides, 91405, Orsay, France}
\affiliation{State Key Laboratory for Superlattices and Microstructures, Institute of Semiconductors, Chinese Academy of Sciences, Beijing 100083, China} 
\affiliation{College of Materials Science and Opto-Electronic Technology, University of Chinese Academy of Sciences, Beijing 100083, China}

\author{Steffi Y. Woo}
\affiliation{Universit\'e Paris-Saclay, CNRS, Laboratoire de Physique des Solides, 91405, Orsay, France}

\author{Nianjheng Wu}
\affiliation{Universit\'e Paris-Saclay, CNRS, Laboratoire de Physique des Solides, 91405, Orsay, France}
\affiliation{Universit\'e Paris-Saclay, Institut des Sciences Mol\'eculaires d'Orsay, 91405, Orsay, France}

\author{Robert Schneider}
\affiliation{Institute of Physics and Center for Nanotechnology, University of Münster, Wilhelm-Klemm-Straße 10, 48149 Münster, Germany}

\author{Andrew J. Mayne}
\affiliation{Universit\'e Paris-Saclay, Institut des Sciences Mol\'eculaires d'Orsay, 91405, Orsay, France}

\author{Steffen Michaelis de Vasconcellos}
\affiliation{Institute of Physics and Center for Nanotechnology, University of Münster, Wilhelm-Klemm-Straße 10, 48149 Münster, Germany}

\author{Ashish Arora}
\affiliation{Institute of Physics and Center for Nanotechnology, University of Münster, Wilhelm-Klemm-Straße 10, 48149 Münster, Germany}

\author{Benjamin J. Carey}
\affiliation{Institute of Physics and Center for Nanotechnology, University of Münster, Wilhelm-Klemm-Straße 10, 48149 Münster, Germany}

\author{Johann A. Preuß}
\affiliation{Institute of Physics and Center for Nanotechnology, University of Münster, Wilhelm-Klemm-Straße 10, 48149 Münster, Germany}

\author{No\'emie Bonnet}
\affiliation{Universit\'e Paris-Saclay, CNRS, Laboratoire de Physique des Solides, 91405, Orsay, France}

\author{Cecilia Mattevi}
\affiliation{Department of Materials, Imperial College London, London SW7 2AZ, UK}

\author{Kenji Watanabe}
\affiliation{Research Center for Functional Materials, National Institute for Materials Science, 1-1 Namiki, Tsukuba 305-0044,Japan}

\author{Takashi Taniguchi}
\affiliation{International Center for Materials Nanoarchitectonics, National Institute for Materials Science, 1-1 Namiki, Tsukuba 305-0044,Japan}

\author{Zhichuan Niu}
\affiliation{State Key Laboratory for Superlattices and Microstructures, Institute of Semiconductors, Chinese Academy of Sciences, Beijing 100083, China} 
\affiliation{College of Materials Science and Opto-Electronic Technology, University of Chinese Academy of Sciences, Beijing 100083, China}

\author{Rudolf Bratschitsch}
\affiliation{Institute of Physics and Center for Nanotechnology, University of Münster, Wilhelm-Klemm-Straße 10, 48149 Münster, Germany}

\author{Luiz~H.~G.~Tizei}
\email{luiz.galvao-tizei@universite-paris-saclay.fr}
\affiliation{Universit\'e Paris-Saclay, CNRS, Laboratoire de Physique des Solides, 91405, Orsay, France}

\begin{abstract}
The excitonic states of transition metal dichacolgenide (TMD) monolayers are heavily influenced by their external dielectric environment based on the substrate used. In this work, various wide bandgap dielectric materials, namely hexagonal boron nitride (\textit{h}-BN) and amorphous silicon nitride (Si$_3$N$_4$), under different configurations as support or encapsulation material for WS$_2$ monolayers are investigated to disentangle the factors contributing to inhomogeneous broadening of exciton absorption lines in TMDs using electron energy loss spectroscopy (EELS) in a scanning transmission electron microscope (STEM). In addition, monolayer roughness in each configuration was determined from tilt series of electron diffraction patterns by assessing the broadening of diffraction spots by comparison with simulations. From our experiments, the main factors that play a role in linewidth broadening can be classified in increasing order of importance by: monolayer roughness, surface cleanliness, and substrate-induced charge trapping. Furthermore, because high-energy electrons are used as a probe, electron beam-induced damage on bare TMD monolayer is also revealed to be responsible for irreversible linewidth increases. \textit{h}-BN not only provides clean surfaces of TMD monolayer, and minimal charge disorder, but can also protect the TMD from irradiation damage. This work provides a better understanding of the mechanisms by which \textit{h}-BN remains, to date, the most compatible material for 2D material encapsulation, facilitating the realization of intrinsic material properties to their full potential. 
\end{abstract}
\maketitle

\section{Introduction}

Two-dimensional materials have attracted the interest of a large portion of the solid state physics and nanoscience communities since the production of atomically thin graphene layers \cite{Geim2010}. This was in great part due to the novel physical properties which became accessible in these van der Waals materials as a function of thickness. Among the large class of such materials \cite{Mounet2018}, transition metal dichalcogenide (TMD) monolayers specifically have been widely studied. A monolayer of these materials with the form MX$_2$ (M~=~W or Mo and X~=~S or Se) and the 2H phase are direct bandgap semiconductors \cite{Mak2010,splendiani_emerging_2010,mueller_exciton_2018,wang_colloquium_2018}. More interestingly, the exciton binding energy in them is large (hundreds of meV) \cite{Ugeda2014,arora_magneto-optics_2021}, allowing excitons to survive up to high temperatures. The lower energy excitons are formed by bands close to the \(K\) and \(K^{\prime}\) points. The \textit{d} orbital character from the transition metal of these bands leads to strong spin-orbit coupling, which forms two distinct excitons, denoted X$_A$ and X$_B$ here, with large energy separation (from 80 to 400 meV) \cite{Xiao2012}. Furthermore, the spin-valley states created by the combination of the crystal structure and spin-orbit interaction can be manipulated by polarized light beams \cite{Xiao2012,zeng_valley_2012,sallen_robust_2012}.

Because of the decreased screening of the Coulomb interaction in 2D materials, their excitonic physics is different from that in 3D \cite{Thygesen2017,cudazzo_dielectric_2011}, with deviations from the Rydberg series for 3D materials \cite{Chernikov2014,Thygesen2017,he_tightly_2014}. Also due to the reduced screening, excitons in these materials are quite sensitive to the local dielectric environment \cite{Fang2019,Lorchat2020} or strain \cite{Dominik2020}, for example. Flat and clean samples are the key to preserve the intrinsic high optical quality of TMD in practice \cite{Cadiz2017,Bonnet2021}. This comes with the penalty that reproducible devices are difficult to fabricate, as their optical response is heavily influenced by disorder in their environment \cite{Rhodes2019}. Moreover, suspended monolayers like graphene and MoS$_2$ have an intrinsic rippling nature \cite{dai2017surface,locatelli2010corrugation,Meyer2007, Thomsen2017,kirilenko2011measuring}. For example, low energy electron diffraction (LEED) has demonstrated that suspended 2H-MoS$_2$ monolayer has surface corrugation that decreases with increasing number of layers \cite{dai2017surface}. It has been observed that encapsulation in \textit{h}-BN significantly improves the properties of TMD monolayers, resulting in sharper emission linewidths, approaching the homogeneous broadening limit \cite{Cadiz2017,wierzbowski_direct_2017,ajayi_approaching_2017,Shree2021}. The exact reasons why \textit{h}-BN encapsulation improves the optical properties of TMDs are not completely known. Atomic flatness of \textit{h}-BN is observed to play a role in several transport experiments of graphene encapsulated in \textit{h}-BN \cite{pezzini2020high,barrios2017electrical,vallejo_bustamante_detection_2021}. Furthermore, electron diffraction experiments have shown that \textit{h}-BN encapsulated graphene \cite{Meyer2007, Thomsen2017,kirilenko2011measuring} has a much smaller mean roughness (12 $\pm$ 5 pm) than freestanding monolayer (114 $\pm$ 1 pm) \cite{Thomsen2017}. The observed roughness of encapsulated graphene is comparable to the 27 pm found in scanning tunneling microscopy (STM) and grazing incidence fast atom diffraction measurements on epitaxial graphene on SiC \cite{zugarramurdi_determination_2015,yang_quantum_2010}. Another clear benefit of encapsulation is the movement of surface residue and subsequent gathering into bubbles, which ensures that clean areas are available \cite{Haigh2012}. 

Optical techniques (optical absorption or photoluminescence (PL)) are typically used to study TMDs. Spectroscopic techniques using the electron beam could provide much higher spatial resolution to study optical behavior \cite{Abajo2010,Kociak2017,Polman2019}, which could provide greater understanding about the role of substrates and residue, when coupled to atomically-resolved imaging techniques \cite{Bonnet2021}. Electron energy loss spectroscopy (EELS) has not been widely used to probe the optical properties of TMD monolayers, with only a handful of reports in the literature \cite{Tizei2015, Tizei2016, Nerl2017, Gogoi2019, Moynihan2020, Hong2021, Susarla2021}. A large part for the lack of interest in this technique was the observation of large absorption linewidths (above 150 meV at full-width at half-maximum (FWHM)), which is considerably larger than one would expect even for room temperature optical measurements. Recently, EELS experiments of \textit{h}-BN encapsulated WS$_2$ have shown that X$_A$ linewidths can be much narrower (FWHM below 30 meV) \cite{Bonnet2021}, approaching values comparable to those measured using optical absorption at similar temperatures around 110~K \cite{Arora2020}. A recent PL study of MoSe$_2$ observed linewidths of 45 meV in PL and 55 meV in STM under ambient conditions at room temperature \cite{pommier_scanning_2019}.

In this work, we explore WS$_2$ monolayers on different substrate configurations (freestanding, \textit{h}-BN, and Si$_3$N$_4$) with the objective to understand the possible effects leading to absorption linewidth broadening (in particular in EELS spectra). Our conclusions show that indeed monolayer roughness and surface residue are crucial to understand the line broadening in TMD monolayers. However, we also show that charge inhomogeneity due to charge trapping can lead to linewidth increase, depending on the substrate used. The series of monolayers configurations used allowed us to qualitatively pinpoint which effects are dominant for linewidth broadening. The increasing order of importance on linewidth is: TMD roughness, surface residue, and substrate charge trapping. Lastly, the effects of electron beam-induced damage to the absorption linewidth of WS$_2$ monolayer on Si$_3$N$_4$ is also demonstrated.

The manuscript is organized into sections describing each of the underlying effects. A description of the experiment and the sample fabrication is given in Section \ref{Methods}. The absorption linewidths for freestanding, 15-nm \textit{h}-BN supported, 5-nm and 15-nm \textit{h}-BN encapsulated, 15-nm Si$_3$N$_4$ supported, and 15-nm Si$_3$N$_4$/50-nm \textit{h}-BN encapsulated WS$_2$ monolayers are reported in Section \ref{Linewidth}. In Sections \ref{Roughness}, \ref{Charging} and \ref{Damage}, we discuss the role of monolayer roughness, substrate charging, and electron beam-induced damage on absorption linewidth broadening. Then, 
a comparison and summary of the effects of the various factors on linewidth, and their ranking of importance are presented in Section \ref{dominant effects}. Finally, the conclusions are presented in Section \ref{Conclusion}.

\section{Methods}
\label{Methods}
Experiments were performed on the ChromaTEM microscope, a modified Nion HERMES 200 equipped with an electron monochromator that allows for energy spreads down to below 10 meV at 60 keV and a side entry stage that can cool the sample down to 110 K. The electron beam energy is set at 60 keV for both EELS and diffraction measurements. 

The convergence and collection half-angles for the EELS measurement are 10 mrad and 21 mrad, respectively. The EELS spectrometer dispersion was 4.46 meV/pixel, recorded onto a Princeton Instruments 
\textbf{KURO} CMOS detector with \(2048\times2048\) pixel array. Two types of spectra are measured in this paper: slow-scan and fast-scan EELS. In order to distinguish beam scan rate and detector accumulating period, sampling dwell time and exposure time are used to describe how long the beam dwells at each scan position before it moving to next position and the time taken to accumulate the spectrum, respectively. Typical exposure time for slow-scan is 150--500 ms per spectrum, with the CMOS detector exposure time the same as the sample dwell time, typical of a scanning/detector scheme for STEM-EELS spectrum imaging. For fast-scan, the beam rasters on top of the sample at a speed of 1 $\mu$s/pixel, such that each spectrum is averaged across the full scanned area (few hundred nm$^2$). All measurements for freestanding WS$_2$ monolayers are made in slow-scan mode, while for Si$_3$N$_4$ supported and Si$_3$N$_4$/\textit{h}-BN encapsulated WS$_2$ monolayers, both types of measurements are performed (see further details in Section \ref{Charging}). The convergence half-angle was set to 1 mrad to approximate an almost parallel beam in TEM for the diffraction measurements. The exposure time is 100 or 300 ms for the diffraction patterns. To reveal the effect of monolayer roughness, tilt series of electron diffraction patterns were acquired, with the sample tilted along the sample holder axis from 0 mrad to 385 mrad ($\sim$22$^{\circ}$) with 35 mrad ($\sim$2$^{\circ}$) step as demonstrated similarly for graphene \cite{Meyer2007, Thomsen2017}.

The samples are made by the viscoelastic stamp method as reported for building van der Waals heterostrutures \cite{castellanos-gomez_deterministic_2014,Purdie2018}. \textit{h}-BN was mechanically exfoliated with Scotch tape from bulk monocrystals. WS$_2$ was either exfoliated with \textbf{Nitto} blue tape or grown by chemical vapor deposition (CVD). 15 nm-thick Si$_3$N$_4$ is provided by a commercial \textbf{PELCO} continuous support film TEM grid. Aside from the Si$_3$N$_4$ series samples, the exfoliated WS$_2$ monolayer sample is transferred onto \textbf{PELCO} holey Si$_3$N$_4$ support film of 200 nm thickness, while all other samples were put on top of commercial Mo-supported \textbf{Quantifoil} holey carbon grids, with more details as shown in Section \ref{Sample} in the supplementary information.

\begin{figure}[H]
\begin{center}
  \includegraphics[width=1\linewidth]{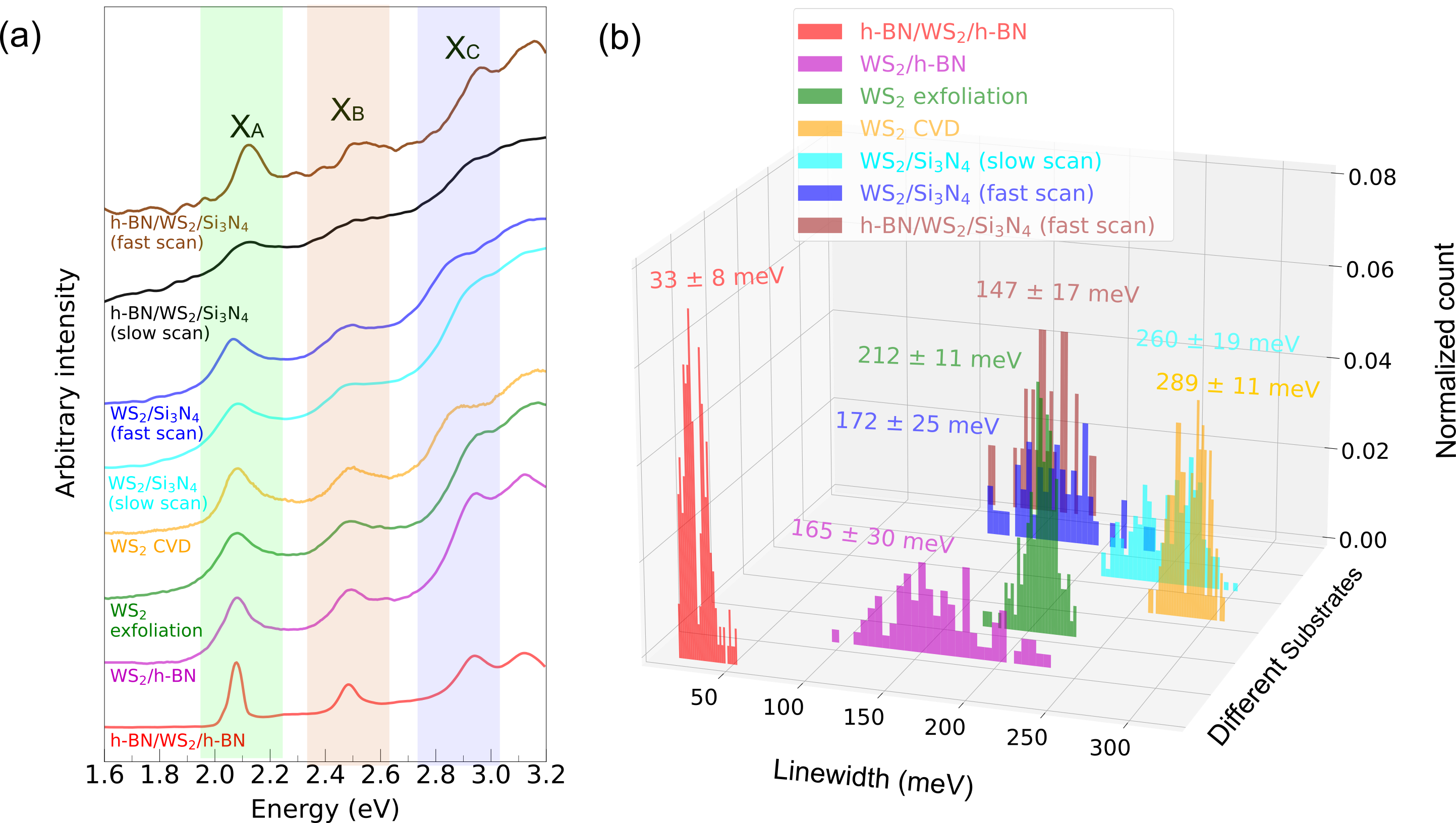}\\
  \captionsetup{font={footnotesize,stretch=1.1},  justification=justified}
  \caption{EELS absorption spectrum and X$_A$ FWHM histogram of monolayer WS$_2$ in various configurations. (a) EELS absorption spectra are measured by STEM-EELS on different substrates at 110~K. EELS spectra are from WS$_2$ monolayers of \textit{h}-BN encapsulated (red), \textit{h}-BN supported (purple), freestanding exfoliated (green), freestanding CVD-grown (yellow), slow-scan Si$_3$N$_4$ supported (cyan), fast-scan Si$_3$N$_4$ supported (blue), slow-scan Si$_3$N$_4$/\textit{h}-BN encapsulated (black) and fast-scan Si$_3$N$_4$/\textit{h}-BN encapsulated (brown) cases, respectively. Each spectrum is normalized with respect to its X$_A$ intensity after zero-loss peak alignment and tail subtraction. (b) Histograms of X$_A$ linewidths measured from different positions of WS$_2$ monolayer on different substrates. Linewidth is extracted from the FWHM of the Lorentzian fit of X$_A$ exciton peak. The mean FWHM and standard deviation for each substrate is listed in the legend. Unfortunately, the excitons in slow-scan EELS spectrum of Si$_3$N$_4$/\textit{h}-BN encapsulated WS$_2$ monolayer is too weak to fit with the strong background of \textit{h}-BN, so there is no histogram for X$_A$ in this configuration.}
  \label{Fig.1}
\end{center}
\end{figure}

\section{Excitonic absorption linewidths on different substrates}
\label{Linewidth}
Exciton linewidth is influenced by several factors, such as the roughness of monolayer, sample cleanliness, and dielectric environment homogeneity. \textit{h}-BN has been shown to be an effective candidate in reducing the absorption and emission linewidths in TMD monolayers \cite{Cadiz2017,ajayi_approaching_2017,wierzbowski_direct_2017}. How does encapsulation with \textit{h}-BN improve the optical quality of TMDs? Here, to answer this question, we explored WS$_2$ monolayers in different configurations: (i) freestanding, (ii) on \textit{h}-BN, (iii) \textit{h}-BN encapsulated, (iv) on Si$_3$N$_4$, and (v) Si$_3$N$_4$/\textit{h}-BN encapsulated.  
In our experiment, three peaks (marked X$_A$, X$_B$ and X$_C$ in Fig. \ref{Fig.1}) can be distinguished in all WS$_2$ monolayers on different substrates. A and B excitons are associated to transitions at the direct band gap at the \(K\) and \(K^{\prime}\) points and are split due to spin-orbit coupling, denoted as X$_A$ and X$_B$. C excitons are associated with direct transitions at the \(Q\) point between \(\Gamma\) and \(K\) point in the Brillouin zone \cite{Qiu2013}. In this paper, we mainly focus on the linewidths of X$_A$ and X$_B$, since they are easily distinguishable and do not overlap. Typical exciton linewidths in various $\sim$micrometer-wide suspended TMD monolayers are estimated above 150 meV at FWHM in EELS \cite{Tizei2015, Nerl2017, Gogoi2019, Hong2021}, both at 150 and 300 K, whereas encapsulated TMD monolayers show in considerably sharper EELS absorption lines, 30 meV at FWHM or below at 150 K \cite{Bonnet2021}, on par with optics \cite{Arora2020,arora_excited-state_2019}.

The absorption EELS spectrum of WS$_2$ monolayer changes for different substrate configurations at 110 K (Fig. \ref{Fig.1}). Small energy shifts occur due to different real parts of the dielectric function of the substrate and strain. More evident is the large variation in width of the absorption peaks, in particular towards the lower-energy transition of X$_A$. This can be quantified by the FWHM measured from spectral fits using Lorentzian functions. As shown in Fig. \ref{Fig.1}(b), the use of \textit{h}-BN induces an obvious improvement in optical quality in comparison to both freestanding and Si$_3$N$_4$ supported monolayers, despite Si$_3$N$_4$ also being a wide-gap dielectric material. For \textit{h}-BN encapsulated WS$_2$ monolayer, the sharpest X$_A$ mean absorption FWHM is observed. This is significantly less than the mean linewidth for the \textit{h}-BN supported WS$_2$ monolayer which exhibits comparable TMD roughness (details for roughness is presented later in Section \ref{Roughness}). It is known that in van der Waals heterostructures, interfaces between materials tend to expel residues, which then coalesce in bubbles \cite{Haigh2012}. Therefore, a monolayer that is only single-side supported has an exposed surface and contains inhomogeneously distributed residue and adsorbates, which would explain the difference in X$_A$ mean FWHM between the \textit{h}-BN supported and encapsulated monolayers. Another supporting evidence is that the \textit{h}-BN and Si$_3$N$_4$ encapsulated WS$_2$ monolayer has sharper linewidth (147 meV) than \textit{h}-BN supported WS$_2$ (165 meV), since the encapsulated monolayer has better surface cleanliness than the supported monolayer. In the next sections, we will explore in greater detail the effect of different substrates to gather more evidence concerning the benefits of \textit{h}-BN encapsulation for TMD monolayers. In short, a combination of factors play a role, while some more than others.

\begin{figure}[H]
\begin{center}
  \includegraphics[width=1\linewidth]{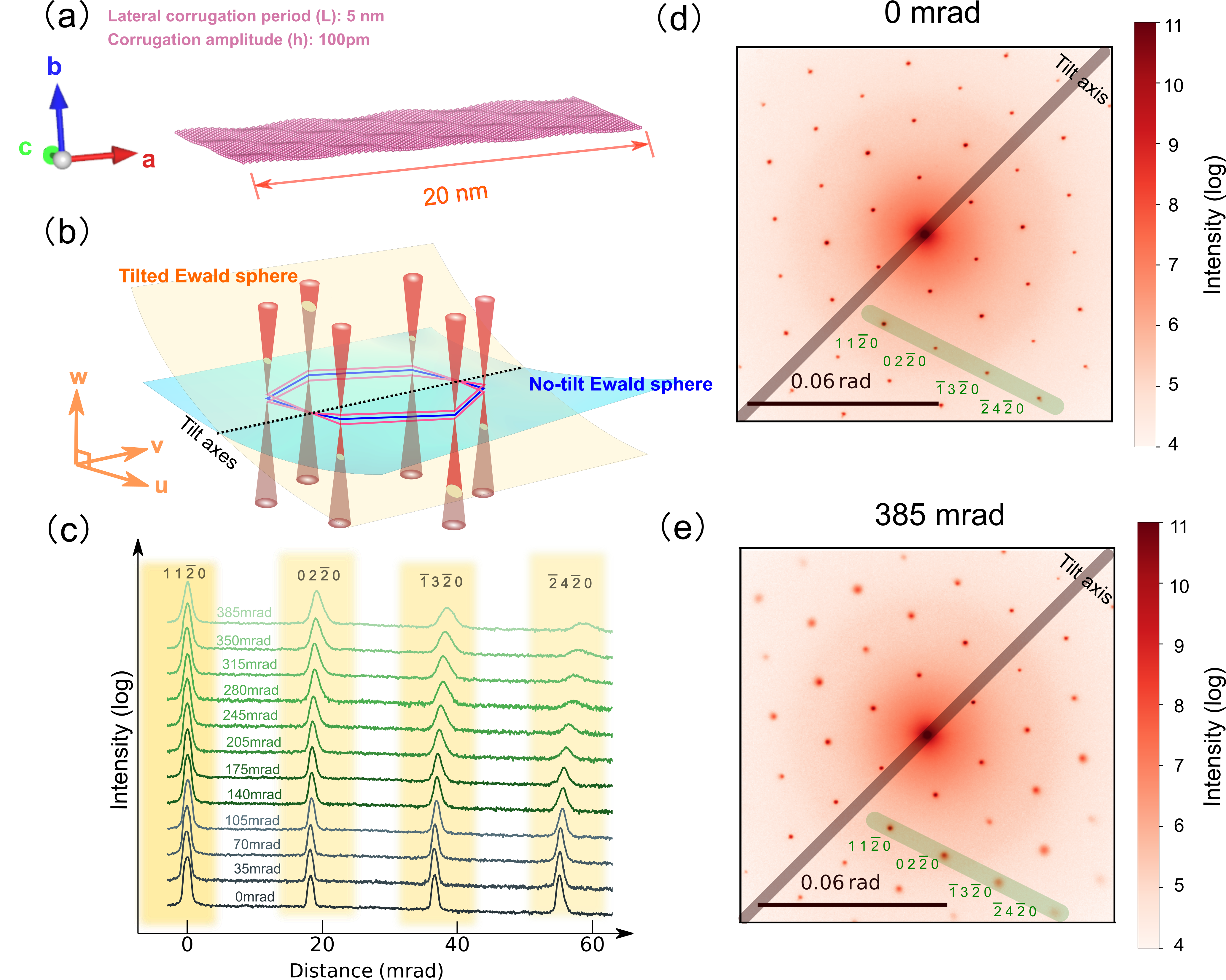}\\
  \captionsetup{font={footnotesize,stretch=1.1},  justification=justified}
  \caption{(a) Model for a rippled WS$_2$ monolayer (only W atoms shown for visibility) displayed using Vesta \cite{momma2011vesta}.  (b) The reciprocal lattice of a rippled TMD monolayer is a set of cones (in red) due to the angular distribution of real-space lattice normal vectors. Also shown are two curved planes which approximate the surface of the Ewald sphere, indicating the points of intersection with the reciprocal space lattice to form the diffraction pattern. The blue plane shows the intersection at zero sample tilt, while the yellow plane shows the intersection at a non-zero tilt angle. The intersection geometry leads to two distinctly different diffraction patterns (d) for zero tilt and (e) for a tilted sample where the spots have become diffuse for a freestanding exfoliated WS$_2$ monolayer. (c) The intensity profiles of the four diffraction spots, which are indicated in (d) and (e) by the green bar, plotted in log scale. The tilt angle is from 0 mrad to 385 mrad with a step size of 35 mrad.}
  \label{Fig.2}
\end{center}
\end{figure}
    
\section{Monolayer roughness on different substrates}
\label{Roughness}
Suspended monolayers have an intrinsic nanoscale corrugation arising from their low-dimensional nature \cite{dai2017surface,locatelli2010corrugation,Meyer2007,Thomsen2017,kirilenko2011measuring}. In addition, sample preparation and transfer protocols to substrates or TEM grids can also result in increased surface corrugation due changes in temperature and surface tension from solvents used. We have observed that freestanding TMD monolayers produced by CVD or mechanical exfoliation have this non-flat fundamental property as illustrated in Fig. \ref{Fig.2}(a). Atomically thin monolayer corrugation can be detected using electron diffraction \cite{Meyer2007, Thomsen2017}, as it can give access to information from the full 3D reciprocal space. In brief, a perfectly flat monolayer should have diffraction spot widths that do not change as the monolayer is tilted with respect to the electron beam. However, a corrugated monolayer would show broadening of diffraction spots as a function of tilt angle as in Fig. \ref{Fig.2}(c).

\begin{figure}[H]
\begin{center}
  \includegraphics[width=1\linewidth]{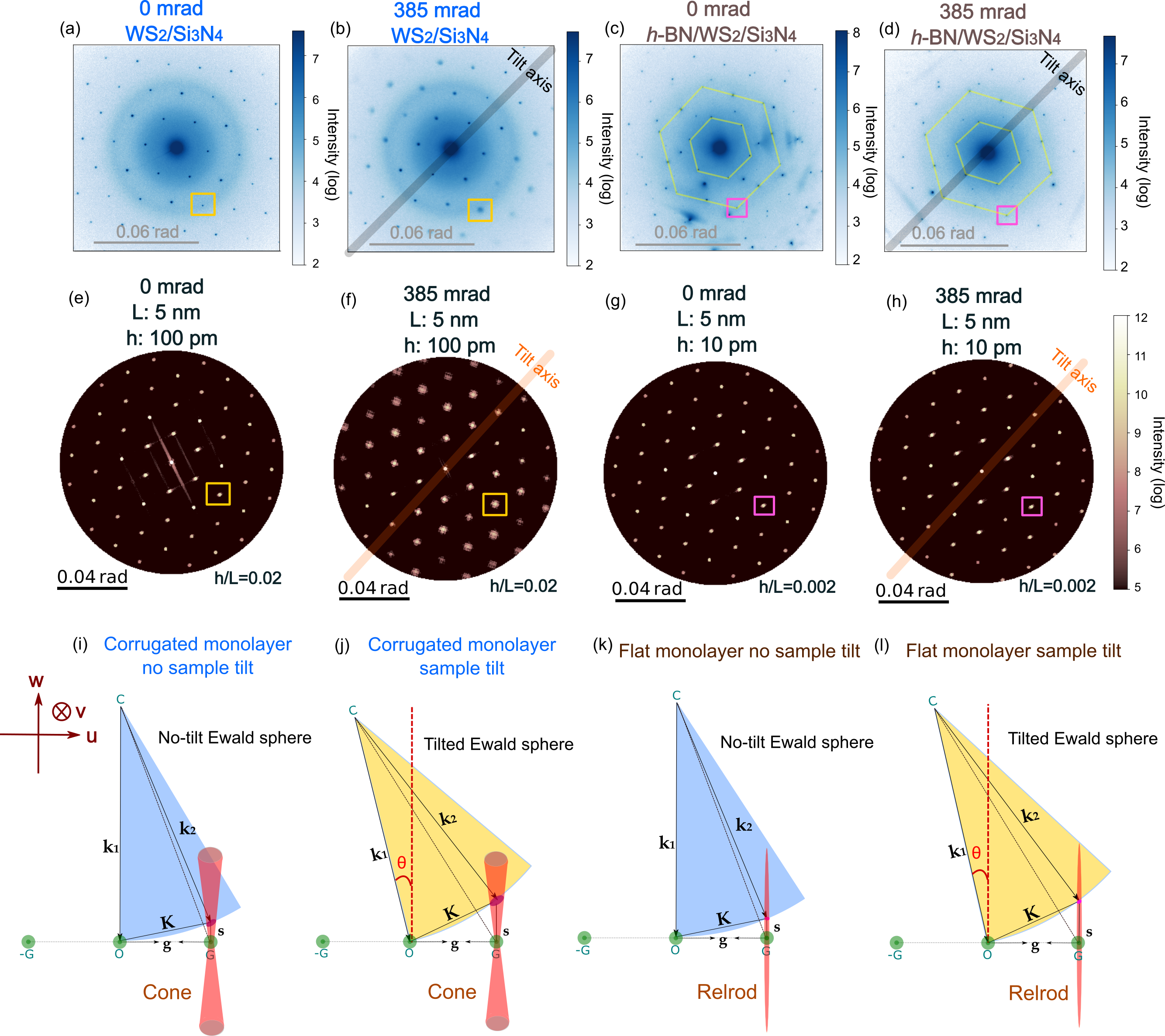}\\
  \captionsetup{font={footnotesize,stretch=1.1},justification=justified}
  \caption{(a) and (b) are experimental diffraction patterns at 0 mrad and 385 mrad for WS$_2$ monolayers supported by Si$_3$N$_4$. (c) and (d) are experimental diffraction patterns at 0 mrad and 385 mrad for WS$_2$ monolayer encapsulated by Si$_3$N$_4$/\textit{h}-BN. (e) and (f) are the simulation results of h/L= 0.02 corrugation at 0 mrad and 385 mrad. Its peak size increased consistently with (a) and (b), as highlighted by the $\langle02\overline{2}0\rangle$ spots boxed in yellow. (g) and (h) are the simulation results of h/L= 0.002 corrugation at 0 mrad and 385 mrad. Its peak size didn't increase, consistent with (c) and (d), also highlighted by the $\langle02\overline{2}0\rangle$ spots boxed in pink. The in-plane rotation and the tilt axis with respect to the first-order reflections in the simulations were chosen arbitrarily, only the tilt axis direction matches with the experiments in this case. (i) and (j) are the reciprocal space of a corrugated monolayer, corresponding to the diffraction patterns in (a), (e) and (b), (f). (k) and (l) are the reciprocal space of a flat monolayer, corresponding the diffraction pattern in (c), (g) and (d), (h). The Ewald sphere is plotted with radius \textit{k} (the difference between incident electron momentum \textbf{k$_1$} and deflected electron momentum \textbf{k$_2$} is very small, so \(k= |\textbf{k$_1$}| \approx |\textbf{k$_2$}|\)) in yellow (tilted sample) and blue (no sample tilt), \(\theta\) in red is the sample tilt angle in (j) and (l). Bragg diffraction condition is relaxed in all atomically thin materials, whether it is flat or corrugated. \(\textbf{K}= \textbf{g}+\textbf{s}= \textbf{k$_2$}-\textbf{k$_1$}\), where the value of \textbf{g} is the inverse of the $\langle02\overline{2}0\rangle$ interplanar distance and \textbf{s} is the vector to describe how far \textbf{K} deviates from the exact Bragg condition.}
  \label{Fig.3}
\end{center}
\end{figure}

To understand this, two points about electron diffraction in thin, corrugated materials should be understood. First of all, because the dimension of the object along the beam propagation direction is small, the reciprocal lattice points along this direction extends more significantly than in thick materials. As shown in Fig. \ref{Fig.3}(k) and (l), the green dots are the reciprocal lattice points of a thick material, whereas they are elongated into reciprocal lattice rods (or relrods \cite{Williams2009}) in red in thin materials. Because they are extended in reciprocal space, the Bragg condition for diffraction is relaxed, meaning that electrons are diffracted even if they are not exactly at the diffraction condition. Thus the Ewald sphere intersects the reciprocal lattice points or relrods with excitation error \textbf{s} = \textbf{K} -- \textbf{g} in this case, where \textbf{g} is the exact Bragg diffraction condition and \textbf{K} is the relaxed Bragg diffraction condition. For this reason, an atomically thin monolayer can be tilted substantially and still show large diffracted intensity and sharp peak width. Moreover, for a corrugated material, instead of relrods perpendicular to the real-space atomic plane, the reciprocal lattice consists of the superposition of relrods due to the real-space lattice normal vectors tilt, leading to the formation of diffuse cones such as the red cones in Fig. \ref{Fig.2}(b) (in \textit{u-v-w} space) and Fig. \ref{Fig.3}(i, j) (in \textit{u-w} plane). The schematic in Fig. \ref{Fig.2}(b) and Fig. \ref{Fig.3}(i, j) illustrates how the Ewald sphere intersects with the reciprocal lattice cones of a rippled monolayer when electrons are incident perpendicularly (zero sample tilt) and non-perpendicularly on the sample (tilted sample). If the sample is not tilted, the Ewald sphere (blue plane in Fig. \ref{Fig.2}(b)) intersects the vertices of the conjoined cones, marked by the blue hexagon in Fig. \ref{Fig.2}(b). Whereas if the sample is tilted, the Ewald sphere (yellow plane in Fig. \ref{Fig.2}(b)) intersects the cones at a certain height along the \textit{w}-axis, denoted by the yellow ovals on the cones. In Fig. \ref{Fig.3}(i) and (l), only one relrod/cone is drawn to show the relaxed Bragg diffraction for flat and corrugated thin material at no sample tilt and tilted sample diffraction configurations.

\begin{figure}[H]
\begin{center}
  \includegraphics[width=0.6\linewidth]{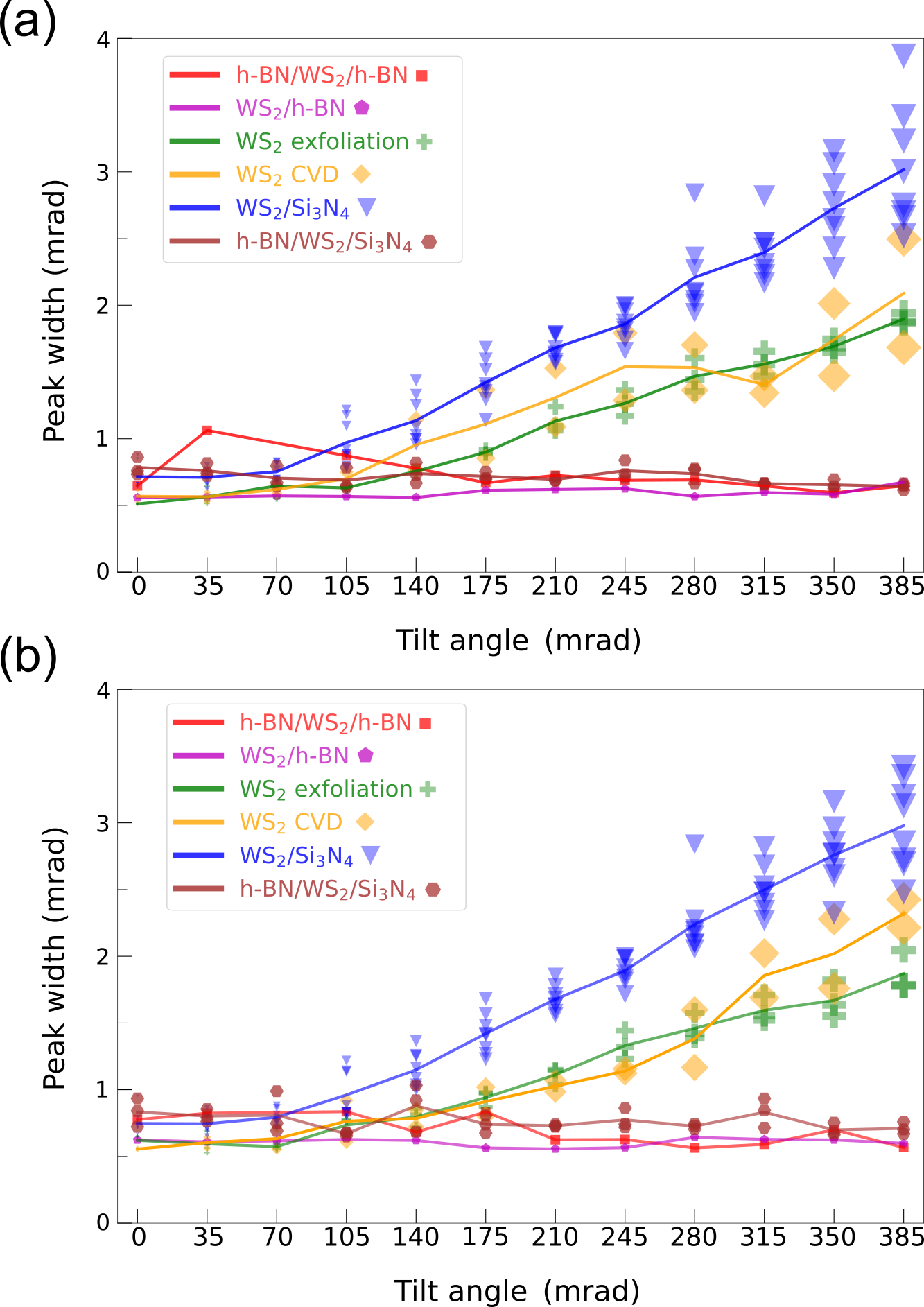}\\
  \captionsetup{font={footnotesize,stretch=1.1},  justification=justified}
  \caption{Roughness measurement of WS$_2$ monolayers for various substrate configurations of: \textit{h}-BN encapsulated (red), \textit{h}-BN supported (purple), freestanding exfoliated (green), freestanding CVD-grown (yellow), Si$_3$N$_4$ supported (blue), and Si$_3$N$_4$/\textit{h}-BN encapsulated (brown) cases. The peak width is extracted from a 2D Gaussian fit of specific diffraction spots in each diffraction pattern, represented by different geometries and colors. The lines indicate the average value of the peak width on different substrates at each angle in the cases of multiple measurements from different regions. (a) FWHM of peak width change with tilt angle along \textit{x} direction. (b) FWHM of peak width change with tilt angle along \textit{y} direction. See the candidate peaks and fit details in Section \ref{Width of diffraction peaks} in the supplementary information.}
  \label{Fig.4}
\end{center}
\end{figure}

Now that we have explained how roughness measurements were performed, we turn to the measurements for different substrates, to try to relate possible variations to the observed linewidth differences seen in the EELS spectra. To compare the roughness of TMD monolayers when placed on or encapsulated in different substrates, we measured the diffraction spot width evolution in a tilt series with steps of 35 mrad (or $\sim$2$^{\circ}$) from 0 mrad up to 385 mrad (or $\sim$22$^{\circ}$), as shown in Fig. \ref{Fig.4}. The width of diffraction spots were retrieved from the FWHM of a 2D Gaussian fit presented in Section \ref{Width of diffraction peaks} in the supplementary information. The 2D Gaussian fit was made at linear intensity scale as they are true values, whereas the plot in Fig. \ref{Fig.2} is shown in log intensity scale to suppress the central spot and enhance weak higher-order spots so that they can be visualized on the same plot.

For the freestanding monolayer and Si$_3$N$_4$ supported monolayer, the width of the diffraction spots quickly increased, with a maximum width when the monolayer is tilted to the maximum angle of 385 mrad, as shown in Fig. \ref{Fig.4} (green, yellow and blue lines). However, the diffraction spot width barely changes as a function of tilt when the monolayer is encapsulated or supported by \textit{h}-BN as in Fig. \ref{Fig.4} (red, purple and brown lines) and Fig. \ref{Fig.3} (c, d). Both a single thin layer of 15-nm \textit{h}-BN (supported monolayer) or full \textit{h}-BN encapsulation reduces roughness, producing a flat WS$_2$ monolayer, as shown in Fig. \ref{Fig.SI5_hBN_Flat}. The roughness of a WS$_2$ monolayer in different configurations from the roughest to flattest is: Si$_3$N$_4$ supported, freestanding, \textit{h}-BN on one or both sides.

To confirm that the  diffraction spot broadening can be explained by the TMD monolayer roughness, we have performed numerical calculations for the expected diffraction pattern of rough and flat monolayers using QSTEM \cite{Koch2002}. In the atomic model, the roughness was simulated by a product of two trigonometric functions, but the real pattern is most probably more complex. This might explain why the symmetry of the spots observed in experiments do not always match the symmetry observed in simulations. Despite not being relaxed, these model structures are sufficient to view the effect of roughness on the diffraction patterns and estimate the amplitude of the monolayer rippling. The results of the simulations indicate that the typical height (h) for the corrugation roughness equals to 0.01 times the lateral corrugation period (\textit{L}), or \(0.01\times L\), for the freestanding monolayer, \(0.02\times L\) for the monolayer on Si$_3$N$_4$, and less than \(0.002\times L\) for the monolayer supported by and encapsulated in \textit{h}-BN, as shown in Fig. \ref{Fig.SI3freestanding} and Fig. \ref{Fig.SI4Si3N4}. Our simulation thus only provides an estimate for the average roughness amplitude; more details are described in the supplementary information.

These measurements indicate that \textit{h}-BN as substrate suppresses the intrinsic roughness of TMD monolayer thus ensure the flatness of a TMD monolayer, showing at least \(20\times\) reduction in roughness even when used in conjunction on top of rough substrates such as Si$_3$N$_4$ (diffraction patterns shown in Fig. \ref{Fig.SI5_hBN_Flat}). This is expected, in view of previous experiments with \textit{h}-BN-encapsulation of graphene and optical experiments of encapsulated TMDs \cite{Dean2010, Wang2012, Stolyarov2015,arora_dark_2020,arora_excited-state_2019}. A WS$_2$ monolayer on top of the Si$_3$N$_4$ is expected to be rough, as this amorphous substrate is not flat. This is confirmed using electron diffraction, as summarized in Fig. \ref{Fig.3}(a, b) and Fig. \ref{Fig.4}. A separate region of the same WS$_2$ CVD monolayer on Si$_3$N$_4$ covered by a thin \textit{h}-BN layer shows a diffraction pattern with similar spot widths to the \textit{h}-BN encapsulated monolayer, at all sample tilts as shown in Fig. \ref{Fig.3}(c, d) and Fig. \ref{Fig.4}. This allows the construction of high-quality devices on top of different substrates, even if the substrate itself is rough (at least with roughness comparable to that of the amorphous Si$_3$N$_4$), to obtain a flat TMD monolayer by the use of single top layer \textit{h}-BN.

\section{Effect of dose rate on charging}
\label{Charging}

The monolayer supported by Si$_3$N$_4$ is not flat, as discussed in the previous section, which explains the large FWHM for the X$_A$ reported in Fig. \ref{Fig.1}. However, if monolayer roughness was the only cause for the linewidth broadening, one would expect to measure sharper X$_A$ absorption lines on the sample which is covered by a thin \textit{h}-BN layer with similarly low monolayer roughness (Fig. \ref{Fig.4} data points in brown). A reduction is observed in Fig. \ref{Fig.1}(b) from 172 meV to 147 meV in FWHM (mean values of the histogram) when comparing the same fast-scan rates. However, the linewidth in FWHM still is significantly larger than observed in the \textit{h}-BN encapsulated sample (33 meV on average).

During the EELS experiments, it was observed that the measured FWHM of the X$_A$ absorption peak for WS$_2$ across Si$_3$N$_4$ depended on how the electron beam was scanned on the samples. We note that the 15-nm thick Si$_3$N$_4$ layer used to support the WS$_2$ monolayer is an insulator, which is known to be prone to charge accumulation. This would lead to a local shift of the exciton line and inhomogeneous broadening, explaining our observations. The important parameters here are the scan speed (how long the beam dwells at a specific pixel before moving to next pixel), spatial sampling (the distance between neighboring pixels), and total scanned area size. To test and control this effect, we acquired spectra with the same total acquisition times and total scanned area size, while the electron beam was scanned at different rates and sampling on the TMD monolayer. 

\begin{figure}[H]
\begin{center}
  \includegraphics[width=\linewidth]{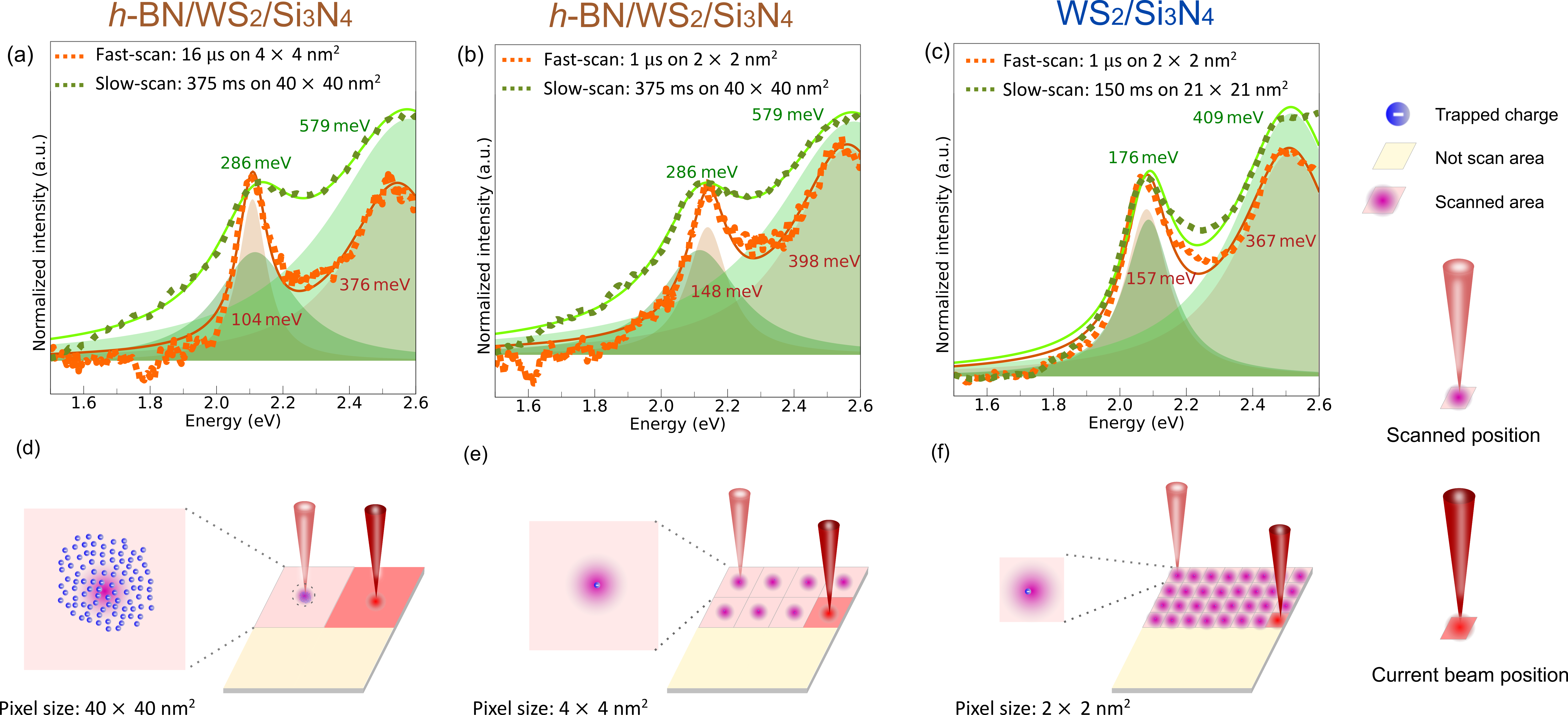}\\
  \captionsetup{font={footnotesize,stretch=1.1},  justification=justified}
  \caption{(a) Slow-scan vs fast-scan in \textit{h}-BN/WS$_2$/Si$_3$N$_4$: spectra comparison between slow-scan (green dotted line) and fast-scan (orange dotted line) acquisition. The pixel dwell times are 375 ms and 16 $\mu$s and the pixel sizes are 40$\times$40 nm$^2$ and 4$\times$4 nm$^2$, respectively. The linewidth is given by the FWHM of a Lorenzian fit of X$_A$ and X$_B$ peaks as labelled, solid line is the sum of the fitted curves of the two excitations and the offset. (b) Slow-scan vs fast-scan (1 $\mu$s on 2$\times$2 nm$^2$) in \textit{h}-BN/WS$_2$/Si$_3$N$_4$:  The linewidth of X$_A$ and X$_B$ in fast-scan (1 $\mu$s on 2$\times$2 nm$^2$) is 148 meV and 398 meV. Linewidth of X$_A$ is wider than fast-scan (16 $\mu$s on 4$\times$4 nm$^2$), but sharper than slow-scan (375 ms on 40$\times$40 nm$^2$ area). (c) Slow-scan vs fast-scan (1 $\mu$s on 2$\times$2 nm$^2$) in WS$_2$/Si$_3$N$_4$: The linewidth of X$_A$ and X$_B$ are as labelled. (d) The scan patterns correspond to slow-scan mode: 100 ms dwell time on a 100 nm$^2$ sized pixel, the most residual charges induced by fast beam. (e) The scan patterns correspond to large pixel scan patterns: $\mu$s dwell time on 4$\times$4 nm$^2$ pixel, the least residual charges induced by fast beam. The linewidth is the closest to intrinsic linewidth. (f) The scan patterns correspond to small pixel scan patterns: 1 $\mu$s dwell time on 2$\times$2 nm$^2$ pixel, less residual charges than slow-scan but more residual charges than large pixel fast-scan.}
  \label{Fig.5}
\end{center}
\end{figure}

Using acquisition settings typical to EELS spectrum imaging (slow-scan), a dwell time of the 100 ms time scale, with an approximately 1-nm wide beam and a sampling of the order of 10 nm/pixel, the X$_A$ was systematically broader than fast-scan as presented in Fig. \ref{Fig.5}(a-c) (green curves). In comparison to spectra acquired over identical regions where the electron beam was rastered at much faster speeds (sample dwell time of 1--16 $\mu$s/pixel) while keeping the exposure time constant, the linewidths can be decreased by more than half (orange curves in Fig.\ref{Fig.5} (a-c)). The spectral evolution due to these changes in scan parameters is summarized schematically in Fig. \ref{Fig.5}(d--f). Globally, typical EELS spectrum imaging acquisition leads to long dwell times for the electron beam, which allows the charges to be trapped (according to our interpretation) in the Si$_3$N$_4$ substrate, represented by little purple dots in Fig. \ref{Fig.5}(d). The pink and red cones represent the electron beam raster start and finish positions, respectively, while the shaded pink and red areas in the specimen plane represent areas scanned by the electron beam and current scan areas. One immediately sees in Fig. \ref{Fig.5}(e,f) that increasing the scan speed and the sampling can distribute these trapped charges more homogeneously across a larger area, reducing their overall effect. 15-nm Si$_3$N$_4$-windowed TEM grids are routinely used for EELS experiments of plasmonic materials, with their charging under the electron irradiation a known limitation, which leads to electron beam displacement with respect to the sample (seen as spatial drift during data acquisition). \textit{h}-BN, while also an insulator on the other hand, has been shown to be a valid alternative, which does not possess this limitation \cite{Tizei2018}.

To obtain the intrinsic exciton linewidth in this sample, we utilised scanning parameters denoted as fast-scan mode, in which the sampling is 4 nm/pixel and the scan speed 16 $\mu$s/pixel. At the same time, with the beam scanning, EELS spectra were acquired using the same detector accumulation time of 100 ms/spectrum as in the slow-scan mode. The X$_A$ linewidth with the slow-scan scheme is 286 meV. The faster scan rate produces sharper X$_A$ linewidth (104 meV). This is reproduced in different areas of the same sample, albeit with small changes in spatial sampling, showing similar behavior in Fig. \ref{Fig.5}(a,b).

To check if the residual X$_A$ absorption broadening was due to the remaining roughness of the WS$_2$ monolayer on Si$_3$N$_4$ a part of this sample was covered with 50-nm thick \textit{h}-BN. Indeed, roughness is reduced by the presence of \textit{h}-BN, as described in the previous section. In these regions, the initial X$_A$ FWHM for slow-scan acquisitions is narrower than in the regions only supported on Si$_3$N$_4$, but the overall reduction from fast-scanning is not more pronounced than in the Si$_3$N$_4$ sample. This indicates that other effects are playing a role. Electron diffraction rules out a difference in roughness, but residual effects due to charge trapping accumulated between measurements over the same area without discharging, despite modifying the scanning parameters, could still increase the linewidth. As we will explain further in the next section, electron beam-induced damage in the Si$_3$N$_4$ supported sample also leads to linewidth broadening.

In this section, we discussed the effect of different scan rates on the X$_A$ linewidth of a WS$_2$ monolayer supported by Si$_3$N$_4$. Interestingly, the same charging effect is not observed on \textit{h}-BN supported or encapsulated monolayers, despite the fact that both materials are insulators and \textit{h}-BN has a larger band gap than Si$_3$N$_4$. Other factors such as the low intrinsic disorder of crystalline h-BN, i.e. low density of atomic defects, makes it a better candidate to counteract against surface charge effects to TMDs. Amorphous Si$_3$N$_4$ on the other hand can exhibit extrinsic charge disorder similarly to SiO$_2$, depending on its defect density and impurities, including hosting trapped charges and surface adsorbates \cite{illarionov_role_2016,lee_comparison_2018}. We note that these experiments were performed with electron current between 1--20 pA in a monochromated electron microscope. Experiments were attempted to control reversibly the observed charge trapping. However, the magnitude of the broadening produced at fixed currents varied at different sample positions. We could not identify underlying reasons for these changes. For this reason, a quantitative assessment of the charge trapping was not possible.

\section{Effect of electron beam damage}
\label{Damage}

A final broadening mechanism that we identified is irradiation damage, i.e. electron beam-induced damage in the case of EELS. 60 keV and higher kinetic energy electrons are known to induce damage on TMD samples either through elastic or inelastic scattering. 
The first mechanism, known as knock-on \cite{Kohl2008, Meyer2012, Mihaila2019}, leads to the removal of atoms. It is more effective at higher kinetic energies \cite{Kohl2008} and should present a cutoff at lower energies, below which knock-on should not be possible, as not enough energy would be available to eject the atom. However, due to the thermal motion of atoms \cite{Meyer2012, Mihaila2019}, knock-on damage can still occur below this threshold. Electronic excitations can also induce below threshold atom removal \cite{Kretschmer2020}. The cutoff is proportional to the atomic mass, as the maximum energy transferred during elastic scattering decreases as the ratio between the mass of the electron and the target atom increases \cite{Kohl2008}. For this reason, knock-on of heavy atoms (compared to carbon, whose cutoff is 86 keV \cite{Smith2001}) should be an ineffective damage mechanism, as in WS$_2$. However, the preexistence of defects can decrease the energy necessary for knock-on, allowing atom removal from edges or voids \cite{kotakoski2012stability}. 

Inelastic scattering leads to the material modification through high energy transfer, which can lead to bond breaking and atomic motion. The cross-section for these events increase at lower kinetic energies for the incident electron beam \cite{Kohl2008} and is thus an effective mechanism for beam damage in TMDs. Typically for TMDs, it has been found that a good compromise for electron microscopy and spectroscopy experiments is to use electron beams with kinetic energy between 60 and 100 keV \cite{Tizei2015, Bonnet2021}. Here we chose to use 60 keV, which also minimizes damage on the \textit{h}-BN encapsulation layer.

In the experiments here and others reported by some of the current authors \cite{Bonnet2021}, in addition to the sharp linewidth of the X$_A$, it was also seen that beam damage is significantly reduced in \textit{h}-BN encapsulated TMD monolayers, allowing even the imaging and spectroscopy of stable monolayer edges. However, this is not true for all substrates. In the Si$_3$N$_4$ supported monolayers, beam damage does occur. This appears as a linewidth broadening as a function of repeated irradiation, which is not reversible by subsequently leaving the measured area unexposed to the electron beam for certain periods. This is in contrast to the mechanism described in the previous section, where trapped charges are recombined, leading to reduction of the X$_A$ linewidth if the electron beam is moved away and then faster scan acquisitions are performed. 

\begin{figure}[H]
\begin{center}
  \includegraphics[width=1\linewidth]{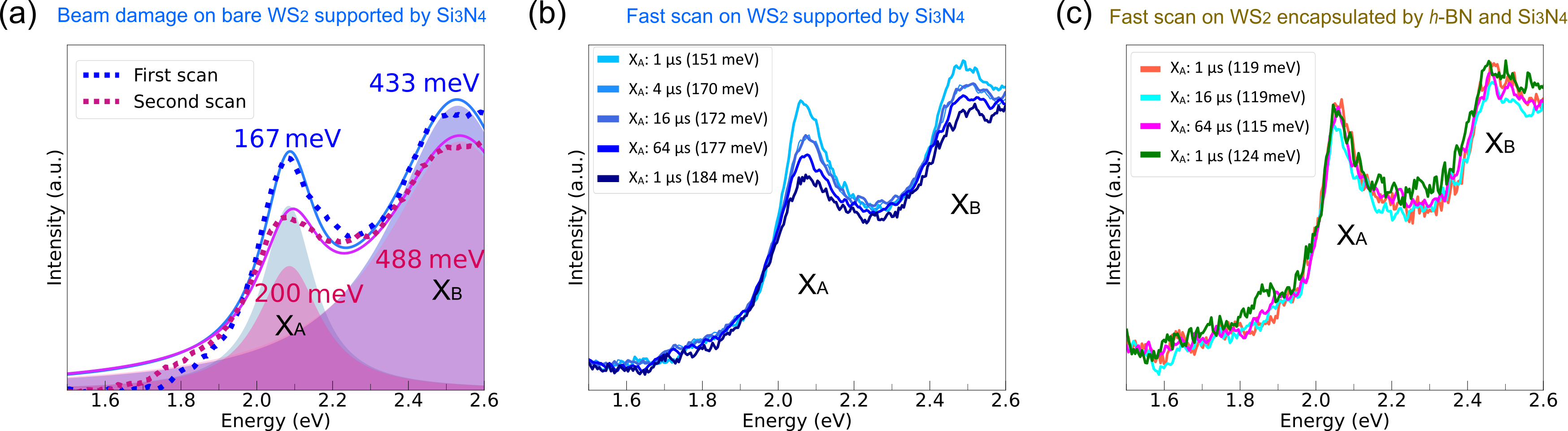}\\
  \captionsetup{font={footnotesize,stretch=1.1},  justification=justified}
  \caption{ (a) WS$_2$ on Si$_3$N$_4$: beam damage induces the broadening of WS$_2$ excitons under EELS slow-scan with pixel size 21$\times$21 nm$^2$ at 150 ms exposure time. (b) WS$_2$ on Si$_3$N$_4$: 2$\times$2 $\mu$m$^2$ area fast-scan with pixel size 2$\times$2 nm$^2$, the area is scanned under the sample dwell time of 1, 4, 16, 64 $\mu$s and then back to 1 $\mu$s. Exciton peak broadening is because of charging and beam damage. (c) The WS$_2$ encapsulated between \textit{h}-BN and Si$_3$N$_4$: 2$\times$2 $\mu$m$^2$ area fast-scan with pixel size 2$\times$2 nm$^2$, the area is scanned under the sample dwell time of 1, 16, 64 $\mu$s and then back to 1 $\mu$s. There is no significant exciton peak broadening after several scans.}
  \label{Fig.6}
\end{center}
\end{figure}

EELS spectra for the same region of WS$_2$ supported on Si$_3$N$_4$ acquired in series show an increase in the X$_A$ linewidth (Fig. \ref{Fig.6}(a)). These spectra are averaged from two \(70\times44\) pixel EELS datacubes with pixel size of 21$\times$21 nm$^2$ and exposure time 150 ms. The effect of the first scan was to induce damage and increase the linewidth from 167 meV to 200 meV. One could argue that this was due to charge trapped in the Si$_3$N$_4$. However, this broadening is not recovered by subsequently leaving the area unexposed to the electron beam. Another test for this hypothesis is the sequential acquisition of spectra in a fast-scan mode (as described in the previous section), with different scan rates. If the induced broadening is reversible, this would indicate that it was due to charging and not permanent crystal damage. This does not occur for the WS$_2$ monolayer on Si$_3$N$_4$. A sequence of acquisitions with fixed scan area and pixel size (2$\times$2 nm$^2$) with sample dwell time 1, 4, 16, 64 and 1 $\mu$s show an increasingly broader X$_A$ absorption peak (from 151 to 184 meV, Fig. \ref{Fig.6}(b)), which is not reduced during the last fast-scan acquisition. The irreversible X$_A$ linewidth increase is not observed for the sample encapsulated in Si$_3$N$_4$/\textit{h}-BN (Fig. \ref{Fig.6}(c)).


\section{Discussion of dominant effects on linewidth}
\label{dominant effects}

\begin{table}[H]
\small
    \centering
    \caption{Summary of (WS$_2$ monolayer) X$_A$ linewidths in different configurations}
    \label{table_summary}
    \begin{tabular}{|c|c|c|c|c|}
\hline
\hline
    \textbf{Configuration}
    &\textbf{Surface cleanliness}
    &\textbf{Trapped charges}
    &\textbf{Roughness}
    &\textbf{Linewidth}\\
\hline    
    \textit{h}-BN/WS$_2$/\textit{h}-BN
    &both sides clean
    &few
    &0.002 $\times$ L
    &33 $\pm$ 8 meV 
    \\
\hline 
     Si$_3$N$_4$/WS$_2$/\textit{h}-BN
     &both sides clean
     &many
     &0.002 $\times$ L
     &\makecell{147 $\pm$ 17 meV \\(fast scan)}
     \\ 
\hline 
     WS$_2$/\textit{h}-BN
     &\makecell{one side clean, \\ one side dirty}
     &few
     &0.002 $\times$ L
     &165 $\pm$ 30 meV
     \\
\hline 
     WS$_2$/Si$_3$N$_4$
     &\makecell{one side clean, \\ one side dirty}
     &many
     &0.02 $\times$ L
     &\makecell{172 $\pm$ 25 meV \\(fast scan)\\
     260 $\pm$ 11 meV \\(slow scan)}
     \\
\hline 
     \makecell{Exfoliated WS$_2$\\(freestanding)}
     &both sides dirty
     &some
     &0.01 $\times$ L
     &212 $\pm$ 11 meV
     \\
\hline 
     \makecell{CVD-grown WS$_2$\\(freestanding)}
     &both sides dirty
     &some
     &0.01 $\times$ L
     &289 $\pm$ 11 meV
\\   
\hline
\hline     
\end{tabular}
\end{table}

In the previous sections we described how various effects influence the EELS linewidth of WS$_2$ excitons in different configurations. By comparing the dominant effect from each substrate and the resulting linewidths (Table \ref{table_summary}) we can reach a qualitative classification of the importance of TMD roughness, TMD surface cleanliness, and substrate charge trapping on linewidth broadening. The comparable roughness (corrugation amplitude, h, as a function of corrugation period, L) for \textit{h}-BN encapsulated, \textit{h}-BN supported and \textit{h}-BN/Si$_3$N$_4$ encapsulated WS$_2$ shows that the monolayer flatness is not the prevailing factor among those considered here on linewidth. For instance, the flatter \textit{h}-BN supported WS$_2$ shows linewidths (165 meV) comparable to that of the rougher Si$_3$N$_4$ supported WS$_2$ configurations (172 meV).

In fact, both surface cleanliness and charge disorder have substantial influence on the linewidth of TMD excitons. Only when the TMD is fully protected against surface charge disorder, presence adsorbates and randomly distributed residue (due to clean interfaces) between two \textit{h}-BN flakes can the narrowest linewdth of 33 meV be obtained. Also, there is no variation in the linewidth of the excitons for either fast-scan or slow-scan in \textit{h}-BN encapsulation, which can further supports that \textit{h}-BN is defect-free and therefore insensitive to residual charge from the electron beam. The linewidth broadening in the case of \textit{h}-BN supported WS$_2$ (165 meV) can be attributed mostly to the lack of cleanliness of its remaining exposed surface.

For monolayer with a comparable surface residue (WS$_2$ on Si$_3$N$_4$) charge trapping on the substrate can significantly increase the linewidth, leading to its linewidth being sensitive to electron beam scanning modes. The effect of the trapped charge on the linewidth is most obvious in the slow-scan mode because it is difficult to discharge at large electron doses. This leads to the local conductivity and dielectric variations and thus inhomogeneous linewidth broadening. This charging effect is less pronounced under fast-scan, where linewidth reduced from 260 meV (slow-scan) to 172 meV (fast-scan). In the case where the charge effect is not obvious like fast-scan in WS$_2$ on Si$_3$N$_4$, its linewidth is similar to the linewidth of WS$_2$ on \textit{h}-BN, since they both have one clean surface and the other exposed. With an addition of a top-layer h-BN, the linewidth is reduced to 147 meV, because both surfaces of WS$_2$ are now cleaner. However, the linewidth is still not as narrow as \textit{h}-BN encapsulated WS$_2$, because residual charges is inevitable from the amorphous Si$_3$N$_4$ even in the case of fast-scan.

The linewidth of WS$_2$ on Si$_3$N$_4$ (260 meV) under slow-scan mode is larger than freestanding exfoliated WS$_2$ (212 meV). This comparison should be taken cautiously, as too much surface residue coupled with high roughness (freestanding CVD-grown WS$_2$) can lead to broader absorption lines (289 meV), although WS$_2$ monolayer corrugates more dramatically on Si$_3$N$_4$ than freestanding.

With these comparisons we conclude that the order of importance of these effect is: monolayer roughness, followed by monolayer cleanliness and substrate charge inhomogeneity. It should be noted that the latter two can be correlated, and not so straight-forwardly ranked in relation to the other in all situations.

\section{Conclusion}
\label{Conclusion}

We have explored the role of different substrates on the EELS absorption linewidths of WS$_2$ monolayers. We have identified four main possible broadening factors: i) monolayer roughness, ii) surface cleanliness, iii) charge trapping in substrates, and iv) electron beam-induced damage.

The first three have been considered and discussed in the past for optical spectroscopies extensively \cite{Dean2010,Cadiz2017,man2016protecting}.
The experiments described here give indications that these three effects on TMD absorption linewidth broadening are also applicable in electron spectroscopy, and can be ranked by decreasing order of dominance of substrate charge disorder, followed by surface cleanliness, then monolayer roughness.
These findings conclude that \textit{h}-BN remains, so far, the most suitable substrate for both optical and electron spectroscopies with TMD monolayers and atomically thin layers for multiple reasons. Clearly, surface cleanliness is crucial: encapsulation with \textit{h}-BN on both surfaces is required to confine the surface residue into localized patches, as widely known in the 2D materials community. Moreover, monolayer flatness is influential to reducing X$_A$ linewidth, as shown by our comparison of \textit{h}-BN supported, \textit{h}-BN encapsulated, and Si$_3$N$_4$/\textit{h}-BN encapsulated WS$_2$ monolayer. However, flatness and cleanliness are clearly not enough as WS$_2$ monolayer encapsulated in Si$_3$N$_4$/\textit{h}-BN is flat and sufficiently clean but still has a significantly broader X$_A$ absorption linewidth than \textit{h}-BN encapsulated WS$_2$ monolayer. Charge trapping on substrates also plays a key role, as exemplified by the experiments on Si$_3$N$_4$. We emphasize that these effects are not intrinsic to electron spectroscopy, and broadly applicable in a similar manner for optical spectroscopies. Trapping of charges in \textit{h}-BN seems to be ineffective, indicating why it is such a good substrate for electron spectroscopy. Finally, electron beam-induced damage also plays a considerable role for electron spectroscopy of 2D materials, intrinsically hindered by the use of an electron beam, but this detriment could be ignored when \textit{h}-BN is on top of the TMD monolayer. The combination of these four characteristics explains the large linewidths observed in EELS experiments in the past \cite{Tizei2015, Tizei2016, Nerl2017, Gogoi2019, Moynihan2020, Hong2021, Susarla2021}.

The experiments reported here indicate a viable path for electron spectroscopy experiments with comparable energy resolution to optical measurements in the available temperature and energy ranges. At the time of writing, electron microscopes capable of 1 meV spectral resolution at liquid helium temperatures and with high spatial resolution are still not available. So a direct comparison with the best optical experiments is not yet possible.

\bibliography{reference.bib}

\section*{Acknowledgement}

This project has been funded in part by the National Agency for Research under the program of future investment TEMPOS-CHROMATEM (reference no. ANR-10-EQPX-50) and National Key Technologies R$\&$D Program of China (Grant No. 2018YFA0306100), the MAGMA research grant (ANR-16-CE09-0027), and the JCJC grant SpinE (reference no. ANR-20-CE42-0020) and by the European Union’s Horizon 2020 research and innovation programme under grant agreement No. 823717 (ESTEEM3) and 101017720 (EBEAM). This work has been supported by Region \^Ile-de-France in the framework of DIM SIRTEQ. 

F.S. acknowledge support by the China Scholarship Council. A.A. and B.J.C. acknowledge support by the Alexander von Humboldt Foundation. C.M. would like to acknowledge the award of a Royal Society University Research Fellowship (UF160539) and the Research Fellow Enhancement Award 2017 (RGF$\backslash$EA$\backslash$180090) by the Royal Society UK. K.W. and T.T. acknowledge support from the Elemental Strategy Initiative conducted by the MEXT, Japan (Grant Number JPMXP0112101001) and JSPS KAKENHI (Grant Numbers 19H05790, 20H00354 and 21H05233).

\newpage
	\setcounter{page}{1}
	\renewcommand\thefigure{SI\arabic{figure}}
	\renewcommand{\thesection}{S\arabic{section}}
	\setcounter{figure}{0} 
	\section*{Supplementary information to Substrate influence on transition metal dichalcogenide monolayer exciton absorption linewidth broadening}
\setcounter{section}{0}	

\section{Sample preparation}
\label{Sample}

\begin{figure}[H]
\begin{center}
  \includegraphics[width=1\linewidth]{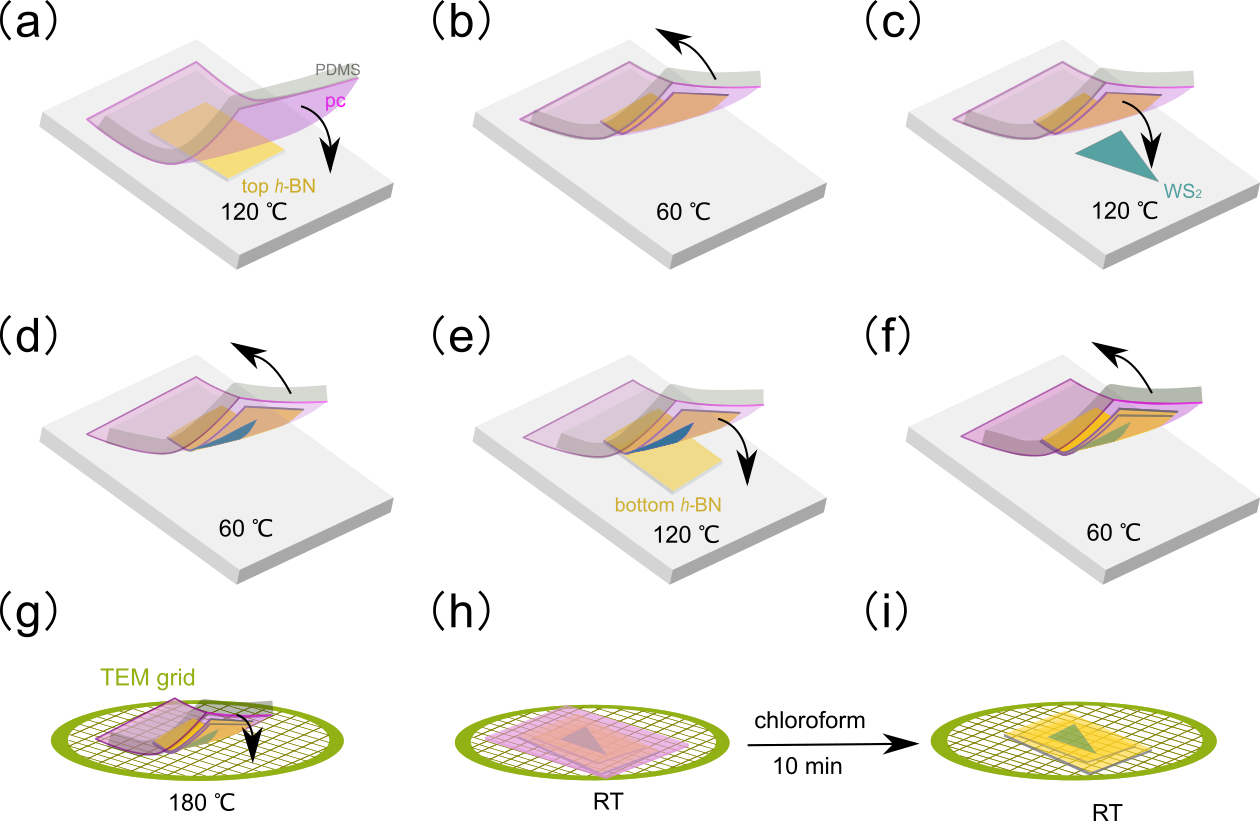}\\
  \captionsetup{font={footnotesize,stretch=1.1},  justification=justified}
  \caption{Scheme of the \textit{h}-BN encapsulated WS$_2$ sample fabrication. (a) PC film (light purple) is pressed by PDMS (grey) to contact \textit{h}-BN (yellow) on the silicon substrate (white). (b) The \textit{h}-BN flake has been picked up by PC. (c) PC film with \textit{h}-BN is ready to pick up monolayer WS$_2$ (dark green). (d) Monolayer WS$_2$ has been picked up by PC with \textit{h}-BN top layer. (e) \textit{h}-BN with monolayer WS$_2$ is ready to pick up another \textit{h}-BN flake. (f) The bottom layer \textit{h}-BN is picked up from the Si substrate. (g) The whole stack is ready to be dropped on top of TEM grid by PDMS. (h) The stack is dropped on top of TEM grid at 180 $^{\circ}$C and PDMS is withdrawn. (i) PC is dissolved in chloroform and the whole stack is firmly on top the TEM grid.  }
  \label{Fig.SI1}
\end{center}
\end{figure}

The samples are prepared by the viscoelastic polymer stamp method \cite{Purdie2018} as shown in Fig. \ref{Fig.SI1}, where the structure of an \textit{h}-BN encapsulated WS$_2$ monolayer is exemplified. A PDMS (polydimethylsiloxane)/PC (polycarbonate) stamp picked up a top \textit{h}-BN layer, WS$_2$ monolayer and bottom \textit{h}-BN sequentially, and finally dropped on top of a Mo-support \textbf{Quantifoil} TEM grid at high temperature. By submerging it into chloroform, the PC film is dissolved. Aside from h-BN encapsulated WS$_2$, we also made samples of freestanding CVD-grown WS$_2$, exfoliated WS$_2$, \textit{h}-BN supported WS$_2$. For \textit{h}-BN encapsulated and supported WS$_2$, the WS$_2$ monolayer is grown by CVD. For the Si$_3$N$_4$ sample series, we also use PDMS/PC stamp pick up of a top \textit{h}-BN, then pick up CVD-grown WS$_2$ monolayer. In this sample, \textit{h}-BN is only half covering the WS$_2$ monolayer. Then the stack is transferred on top of a commercial \textbf{PELCO} 15-nm Si$_3$N$_4$ support membrane based TEM grid. 

\section{QSTEM Simulation}
\label{QSTEM Simulation}

\begin{figure}[H]
\begin{center}
  \includegraphics[width=0.9\linewidth]{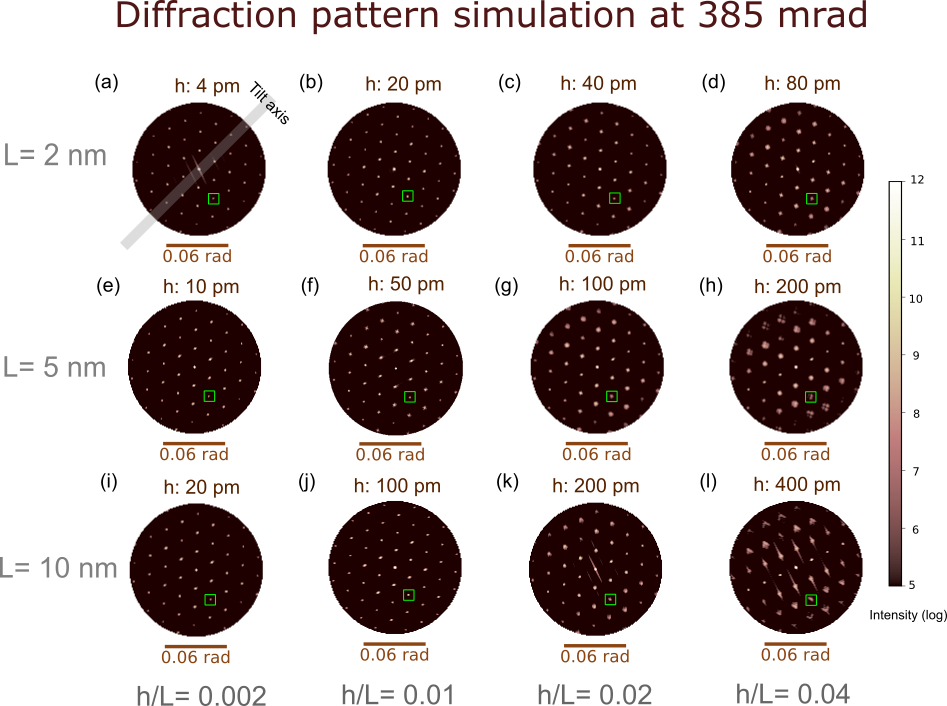}\\
  \captionsetup{font={footnotesize,stretch=1.1},  justification=justified}
  \caption{Diffraction pattern simulation at 385 mrad sample tilt angle of corrugated WS$_2$ monolayer with different ratios between corrugation height (h) and lateral corrugation period (L). h/L = 0.002 in first column (a), (e) and (i); h/L = 0.01 in first column (b), (f) and (j); h/L = 0.02 in first column (c), (g) and (k); h/L = 0.04 in last column (d), (h) and (l). The first row (a)-(d) has the same L= 2 nm; the second row (e)-(h) has the same L= 5 nm; the third row (i)-(l) has the same L= l nm. The tilt axis for all diffraction patterns is denoted in (a) by the grey line.}
  \label{Fig.SI2Simulation}
\end{center}
\end{figure}

The diffraction simulations were calculated using QSTEM \cite{Koch2002}. The beam acceleration voltage and convergence angle is 60 keV and 1 mrad. The astigmatism, spherical aberration and chromatic aberration is 1 nm, 0.01 nm and 1 nm. The brightness is 5 x $10^{8}$ A/$cm^{2}$sr. The window size is 500 \si{\angstrom} with 0.24 \si{\angstrom} resolution for the potential calculation, thus the biggest scattering angle is 66.4 mrad. 500 slices was set to multi-slice the atomic model. Fig. \ref{Fig.SI2Simulation} are the results of the simulations for lateral corrugation period size, L = 2 nm, 5 nm, 10 nm with the h (corrugation amplitude) 0.002, 0.01, 0.02, 0.04 times L.

A model for a rough monolayer was produced by displacing the atoms along the out-of-plane direction (along the \textit{c}-axis, or $[0001]$ direction), while keeping the total projection area of the layer constant. This is achieved by iteratively displacing the atoms along the surface tangent. For better visualization, we plot all experiment and simulation diffraction patterns in log scale. This suppresses the central disk while showing more clearly the weak high-order diffraction peaks. The simulation of diffraction pattern at 0 mrad sample tilt angle is not presented here, because there isn't any difference. The tilt axis is marked in Fig. \ref{Fig.SI2Simulation}(a). This proved that the corrugation is the reason why peaks became larger at high sample tilt angle, comparable to the experimental patterns in Fig. \ref{Fig.2}, Fig. \ref{Fig.4}, and Fig. \ref{Fig.3}. In Fig. \ref{Fig.SI2Simulation}, the diffraction peaks became broaden as the size of h/L increase at 385 mrad sample tilt angle. This means at least in the range of corrugation period of 2--10 nm, it is the h/L ratio instead of discrete L or h values that dictate the width of diffraction peaks at large sample tilt angle, as described by previous reports \cite{Meyer2007,locatelli2010corrugation,dai2017surface}. This is very different from previous studies on freestanding graphene that its diffraction peak size and intensity at high sample tilt angle are governed only by the rippling amplitude \cite{kirilenko2011measuring,Thomsen2017}. 

\begin{figure}[H]
\begin{center}
  \includegraphics[width=0.7\linewidth]{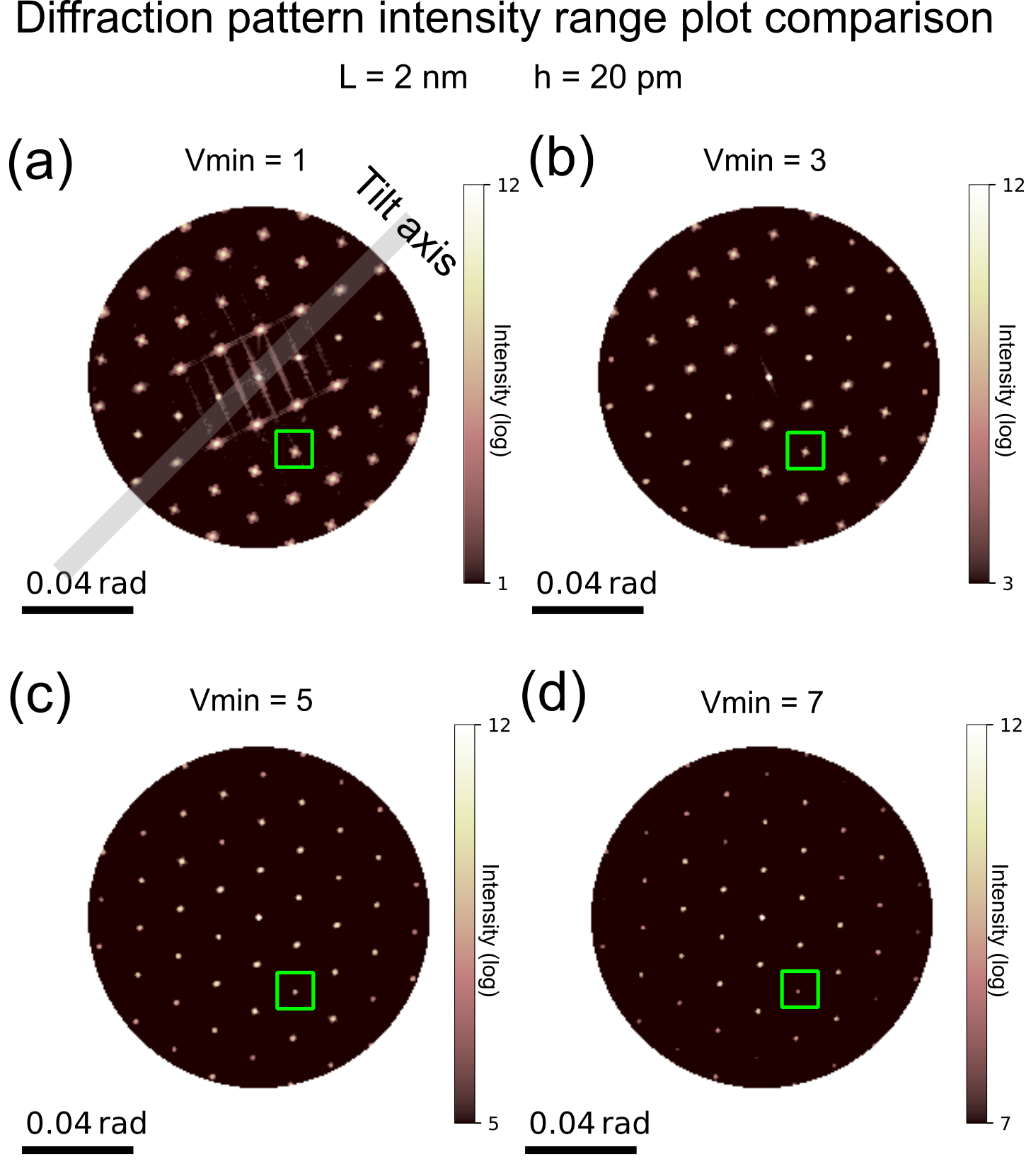}\\
  \captionsetup{font={footnotesize,stretch=1.1},  justification=justified}
  \caption{Different minimum log intensity plot comparison of corrugated WS${_2}$ monolayer simulated diffraction pattern, where L = 2 nm and h = 20 pm. (a), (b), (c) and (d) are same simulation plot with V$_{min}$= 1, 3, 5, 7, respectively.}
  \label{Fig.SI2Simulation_vmin}
\end{center}
\end{figure}

However, there are some artefacts we don't understand when the minimum log intensity for plotting is V$_{min}$= 1 as shown in Fig. \ref{Fig.SI2Simulation_vmin}, especially in Fig. \ref{Fig.SI2Simulation_vmin}(a). It is reasonable because the model use to calculate diffraction pattern has intrinsic long-range periodicity, and cannot be as ideal as the random real world conditions. The atomic model is also not relaxed so some noise also introduced. To suppress this we chose to plot the simulation results at V$_{min}$ = 3 or 5 (minimum log intensity) for the simulated diffraction patterns. This would suppress the artefacts, whereas the peak width at the high-order peaks of weak intensity would a bit smaller, as shown in Fig. \ref{Fig.SI2Simulation_vmin}. In Fig. \ref{Fig.SI2Simulation_vmin}, we plot the simulation of h/L = 0.01 with L = 2 nm corrugated WS${_2}$ monolayer diffraction pattern at 385 sample tilt angle, we do see artefacts at V$_{min}$ = 1 in Fig. \ref{Fig.SI2Simulation_vmin}(a). By plotting it at V$_{min}$ = 3, the noise became weaker. Comparing with others in Fig. \ref{Fig.SI2Simulation_vmin}, the peak size shrinks a bit but this is not so evident in (c) and (b), but peak size shrink too much in (d). Thus, we plotted all simulation results with V$_{min}$ = 3 or 5.

\section{Roughness estimation}
\label{Roughness estimation}

\begin{figure}[H]
\begin{center}
  \includegraphics[width=0.8\linewidth]{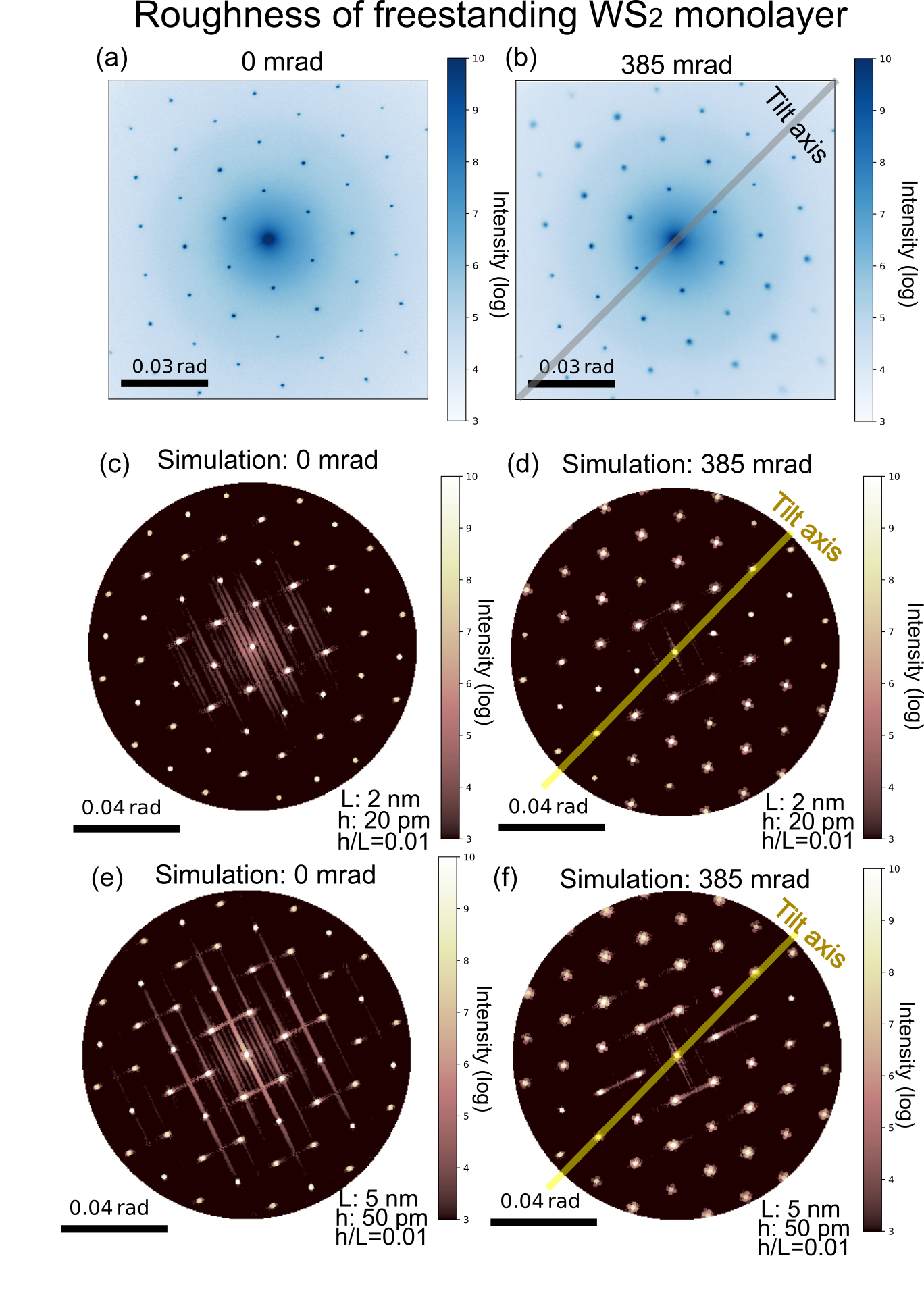}\\
  \captionsetup{font={footnotesize,stretch=1.1},  justification=justified}
  \caption{Roughness of freestanding WS$_2$ monolayer. (a) and (b) are experimental diffraction patterns at 0 and 385 mrad tilt angles. (c) and (d) are the simulation results of h/L = 0.01 with period of L = 2 nm. (e) and (f) are the simulation results of h/L= 0.01 with period of L = 5 nm.}
  \label{Fig.SI3freestanding}
\end{center}
\end{figure}

\begin{figure}[H]
\begin{center}
  \includegraphics[width=0.7\linewidth]{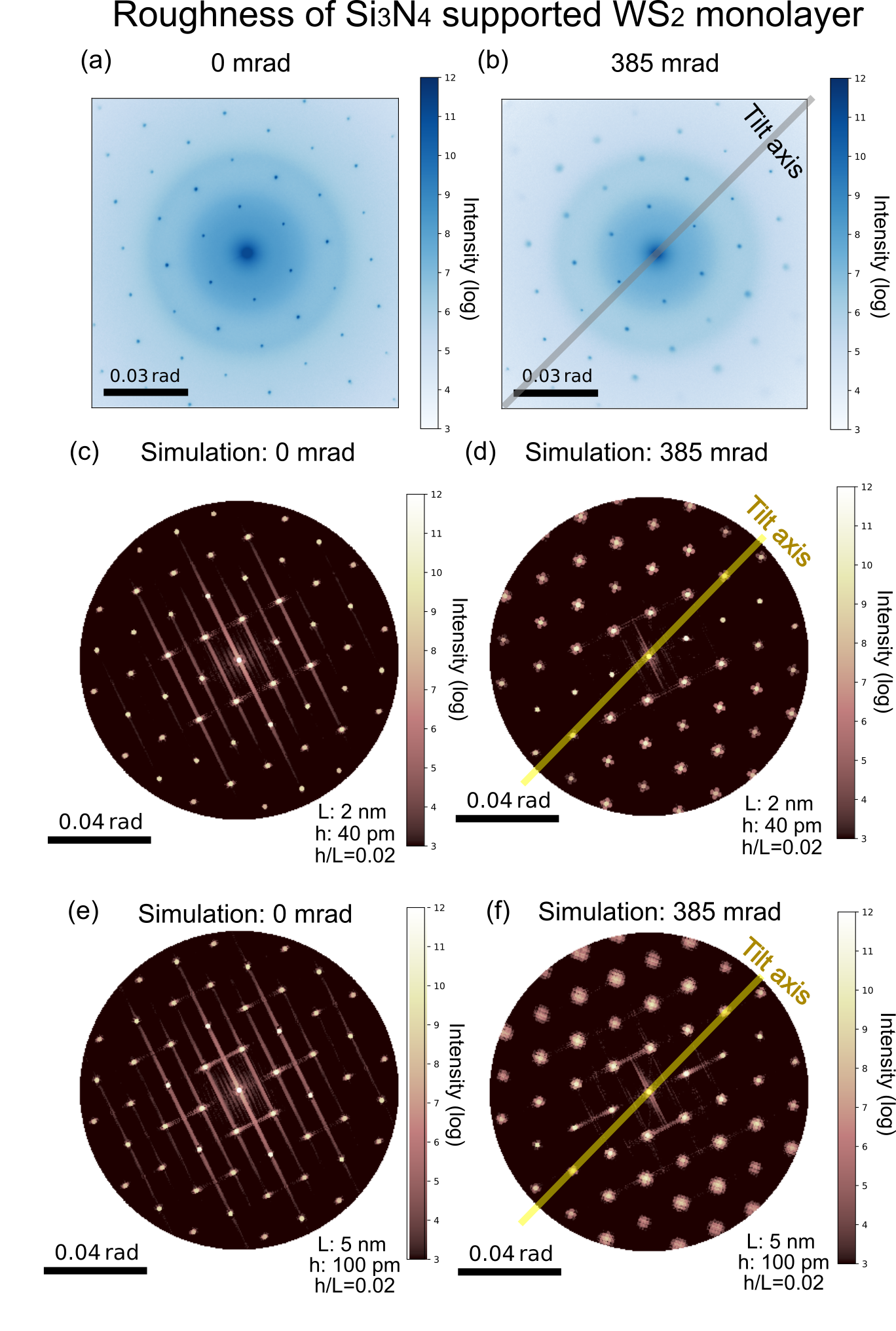}\\
  \captionsetup{font={footnotesize,stretch=1.1},  justification=justified}
  \caption{Roughness of Si$_3$N$_4$ supported WS$_2$ monolayer. (a) and (b) are experimental results of diffraction patterns at 0 and 385 mrad tilt angles. (c) and (d) are the simulation results of h/L= 0.02 with period of L = 2 nm. (e) and (f) are the simulation results of h/L = 0.02 with period of L = 5 nm.}
  \label{Fig.SI4Si3N4}
\end{center}
\end{figure}

\begin{figure}[H]
\begin{center}
  \includegraphics[width=0.7\linewidth]{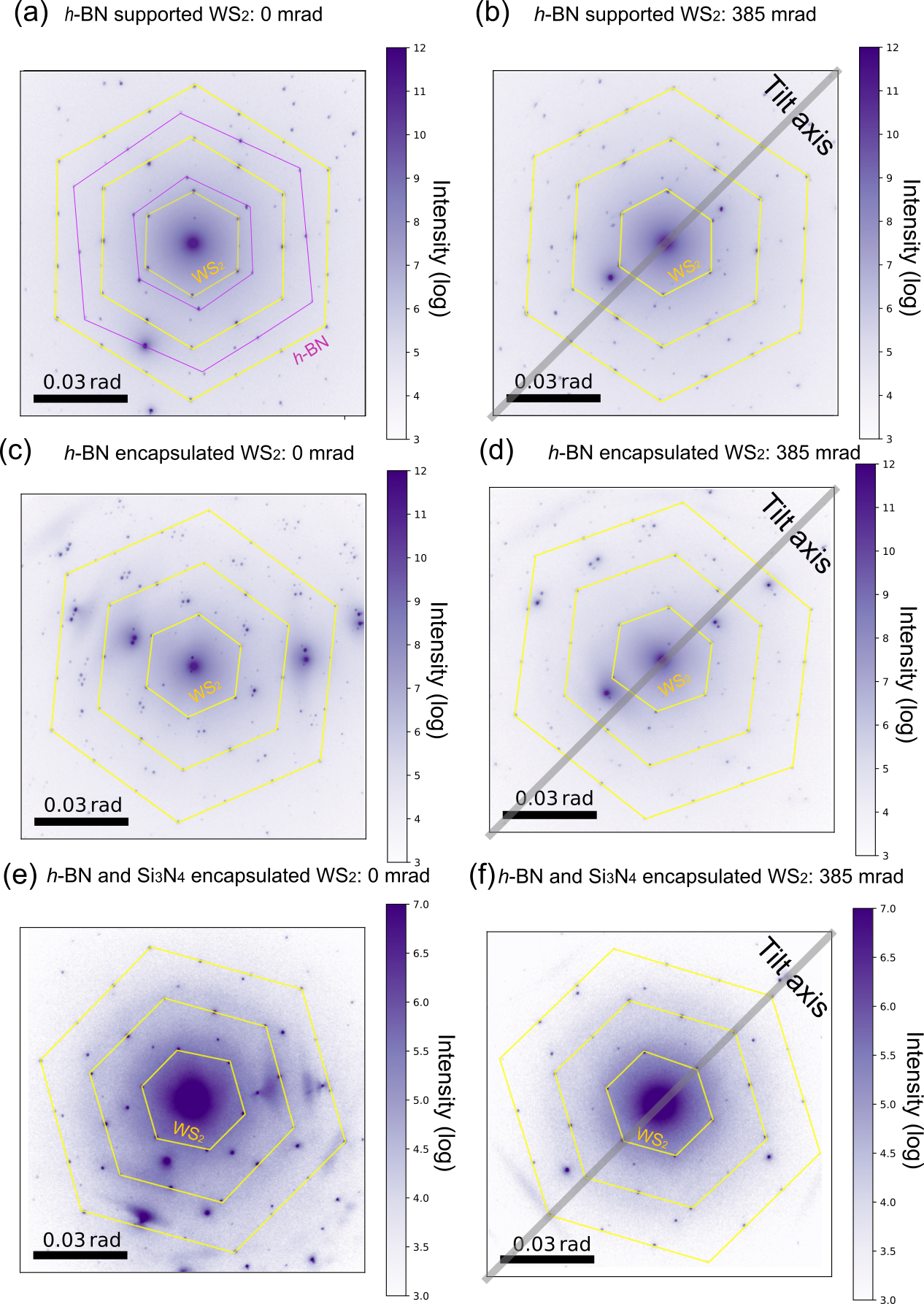}\\
  \captionsetup{font={footnotesize,stretch=1.1},  justification=justified}
  \caption{\textit{h}-BN supported and encapsulated WS$_2$ monolayer diffraction patterns. (a) and (b) are experimental diffraction patterns of \textit{h}-BN supported WS$_2$ monolayer at 0 and 385 mrad. (c) and (d) are experimental diffraction patterns of \textit{h}-BN encapsulated WS$_2$ monolayer at 0 and 385 mrad. (e) and (f) are experimental diffraction pattern of \textit{h}-BN/Si$_3$N$_4$ encapsulated WS$_2$ monolayer at 0 and 385 mrad. The tilt axes are denoted in (b), (d) and (f) by the grey line.}
  \label{Fig.SI5_hBN_Flat}
\end{center}
\end{figure}

In Fig. \ref{Fig.SI3freestanding} and Fig. \ref{Fig.SI4Si3N4}, the experimental data and simulations are plotted in the same manner with V$_{min}$ = 3. In Fig. \ref{Fig.4}, the exfoliated and CVD-grown monolayers have similar results for roughness measurement. Here, we take freestanding exfoliated WS$_2$ as example, the corrugated ratio is around h/L = 0.01. The roughness estimation of Si$_3$N$_4$ supported WS$_2$ is shown in Fig. \ref{Fig.SI4Si3N4}, and its roughness ratio is around h/L = 0.02. Experimental diffraction pattern of \textit{h}-BN on one or both side of WS${_2}$ is shown in Fig. \ref{Fig.SI5_hBN_Flat}. Combined with the simulation in Fig. \ref{Fig.SI2Simulation}(a), (e) and (i), the deduced roughness of samples involving \textit{h}-BN on one or both surfaces is h/L $\leq$ 0.002. 

\section{FWHM of diffraction peaks}
\label{Width of diffraction peaks}

\begin{figure}[H]
\begin{center}
  \includegraphics[width=0.7\linewidth]{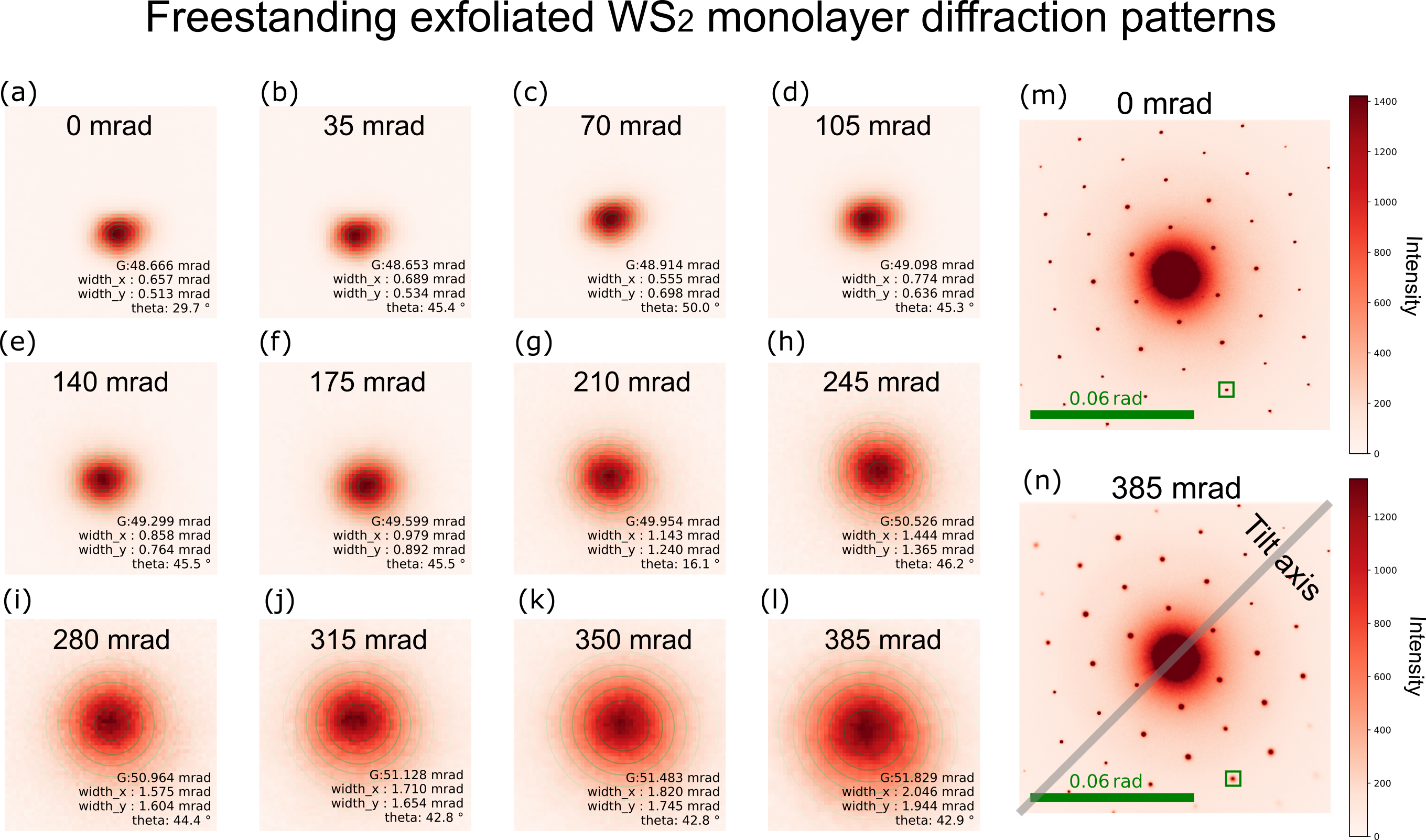}\\
  \captionsetup{font={footnotesize,stretch=1.1},  justification=justified}
  \caption{Two dimensional Gaussian fit of freestanding exfoliated WS$_2$ monolayer diffraction pattern tilt series. (a)-(l) Two dimensional Gaussian fit of the selected peak from 0 mrad to 385 mrad sample tilt angle. The FWHM, rotation angle, and intensity are given at the bottom right corner. The candidate peak selected, $\langle\overline{1}3\overline{2}0\rangle$, is marked by a green rectangle in (m) and (n).}
  \label{Fig.SI6_ex}
\end{center}
\end{figure}

\begin{figure}[H]
\begin{center}
  \includegraphics[width=0.7\linewidth]{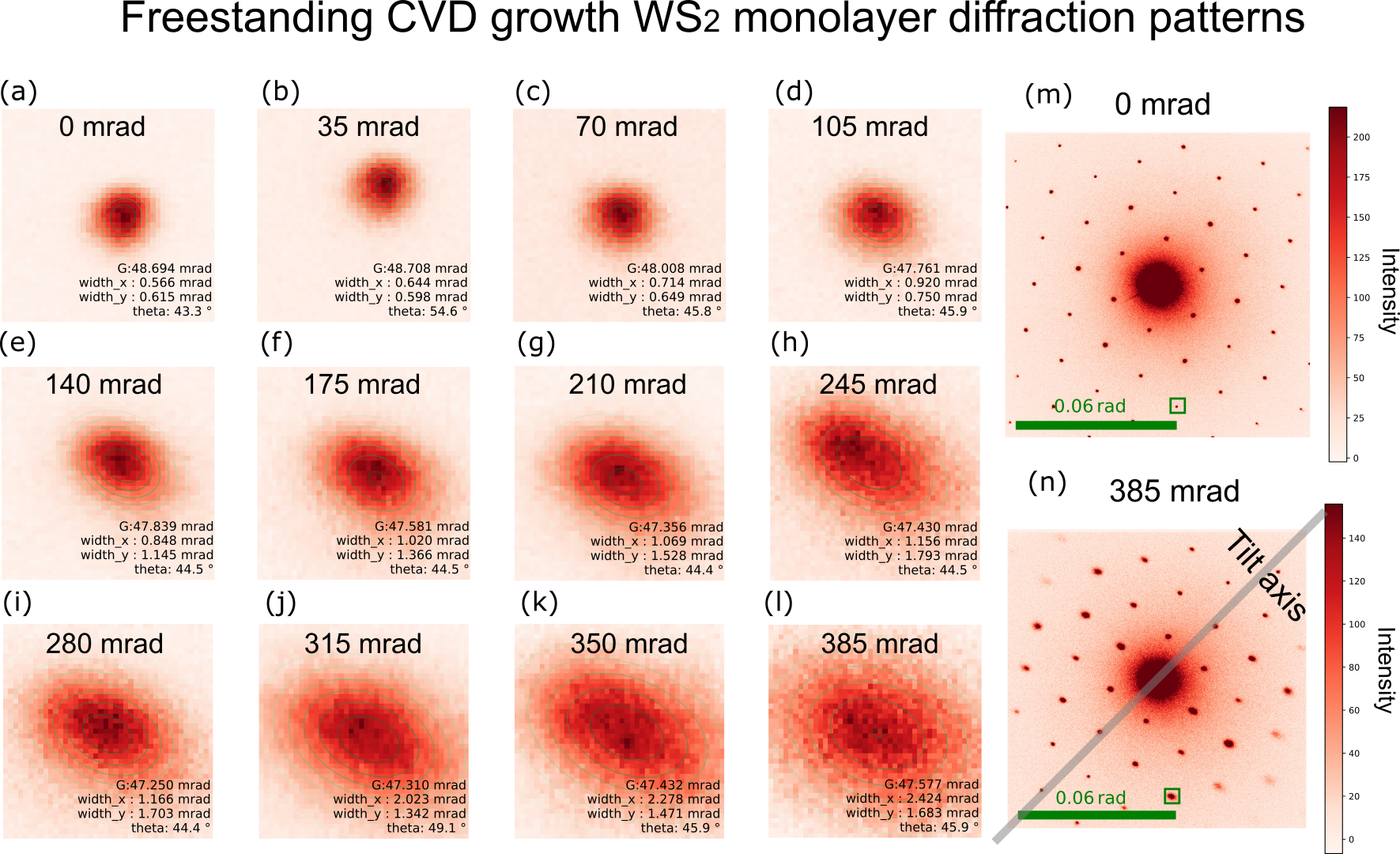}\\
  \captionsetup{font={footnotesize,stretch=1.1},  justification=justified}
  \caption{Two dimensional Gaussian fit of freestanding CVD-grown WS$_2$ monolayer diffraction pattern tilt series. ((a)-(l) Two dimensional Gaussian fit of the selected peak from 0 mrad to 385~mrad sample tilt angle. The FWHM, rotation angle, and intensity are given at the bottom right corner. The candidate peak selected, $\langle\overline{1}3\overline{2}0\rangle$, is marked by a green rectangle in (m) and (n).}
  \label{Fig.SI6_cvd}
\end{center}
\end{figure}

\begin{figure}[H]
\begin{center}
  \includegraphics[width=0.7\linewidth]{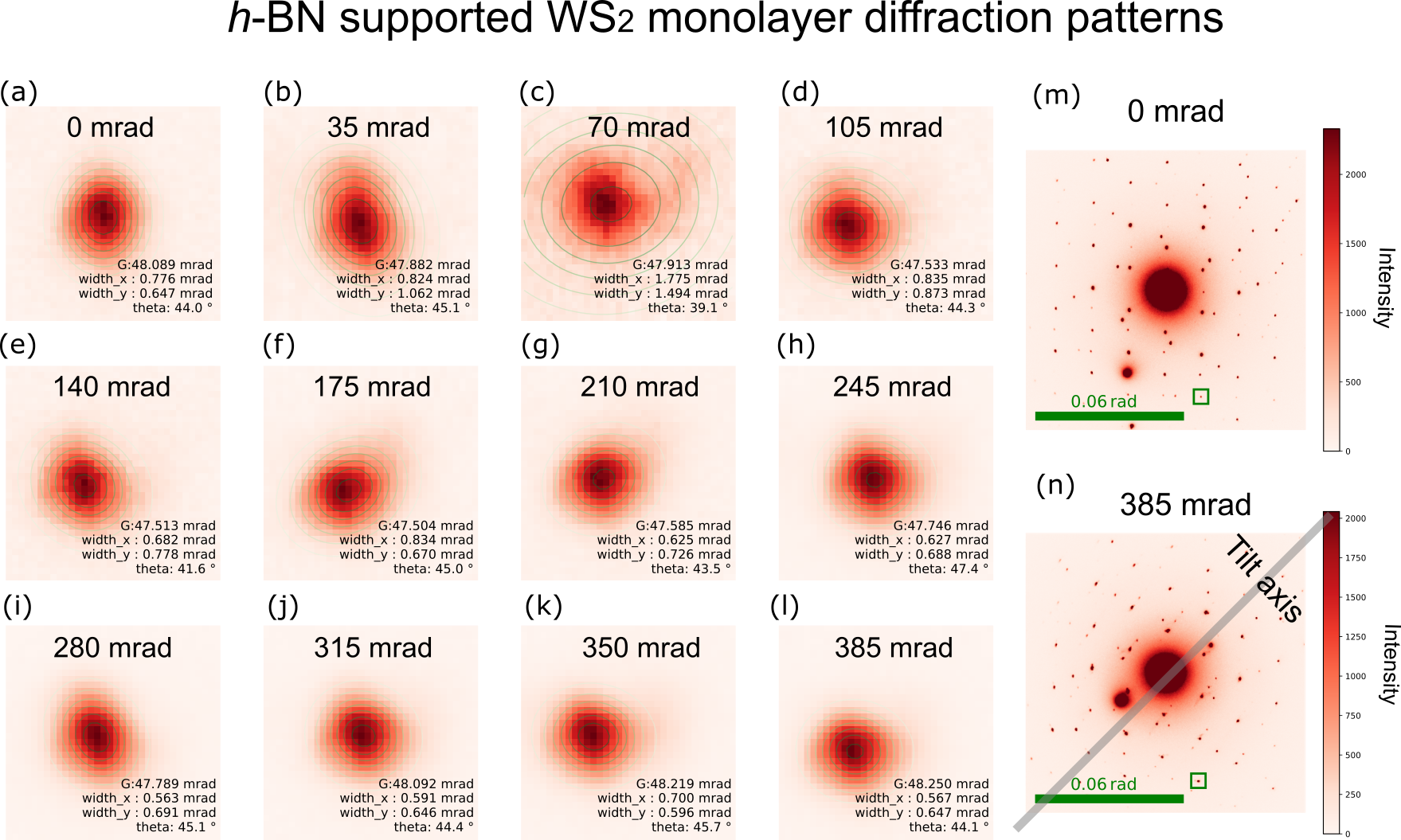}\\
  \captionsetup{font={footnotesize,stretch=1.1},  justification=justified}
  \caption{Two dimensional Gaussian fit of \textit{h}-BN supported WS$_2$ monolayer diffraction pattern tilt series. ((a)-(l) Two dimensional Gaussian fit of the selected peak from 0 mrad to 385~mrad sample tilt angle. The FWHM, rotation angle, and intensity are given at the bottom right corner. The candidate peak selected, $\langle\overline{1}3\overline{2}0\rangle$, is marked by a green rectangle in (m) and (n).}
  \label{Fig.SI6_hBN_S}
\end{center}
\end{figure}

\begin{figure}[H]
\begin{center}
  \includegraphics[width=0.7\linewidth]{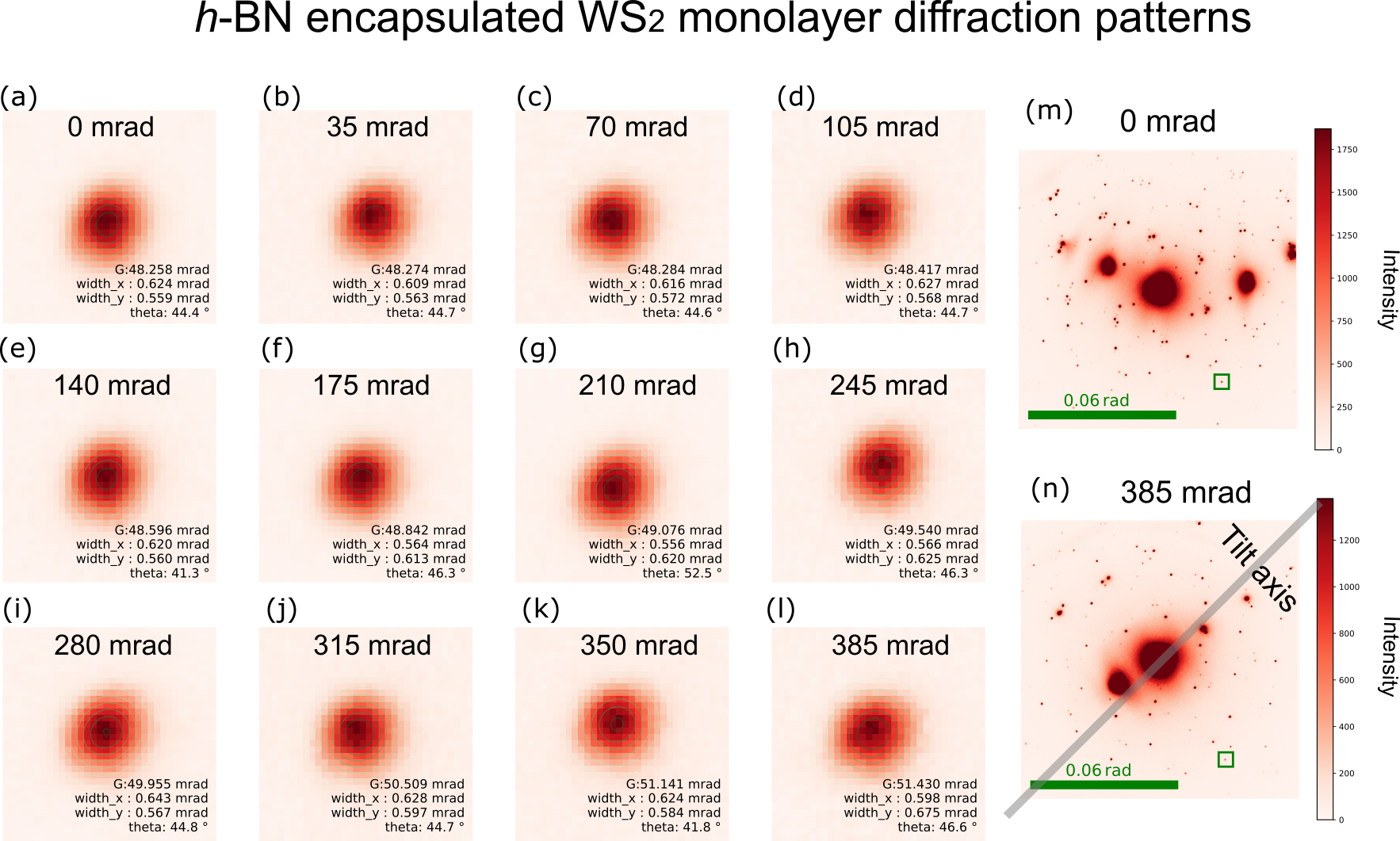}\\
  \captionsetup{font={footnotesize,stretch=1.1},  justification=justified}
  \caption{Two dimensional Gaussian fit of \textit{h}-BN encapsulated WS$_2$ monolayer diffraction pattern tilt series. (a)-(l) Two dimensional Gaussian fit of the selected peak from 0 mrad to 385~mrad sample tilt angle. The FWHM, rotation angle, and intensity are given at the bottom right corner. The candidate peak selected, $\langle\overline{2}3\overline{1}0\rangle$, is marked by a green rectangle in (m) and (n).}
  \label{Fig.SI6_hBN_E}
\end{center}
\end{figure}

\begin{figure}[H]
\begin{center}
  \includegraphics[width=0.7\linewidth]{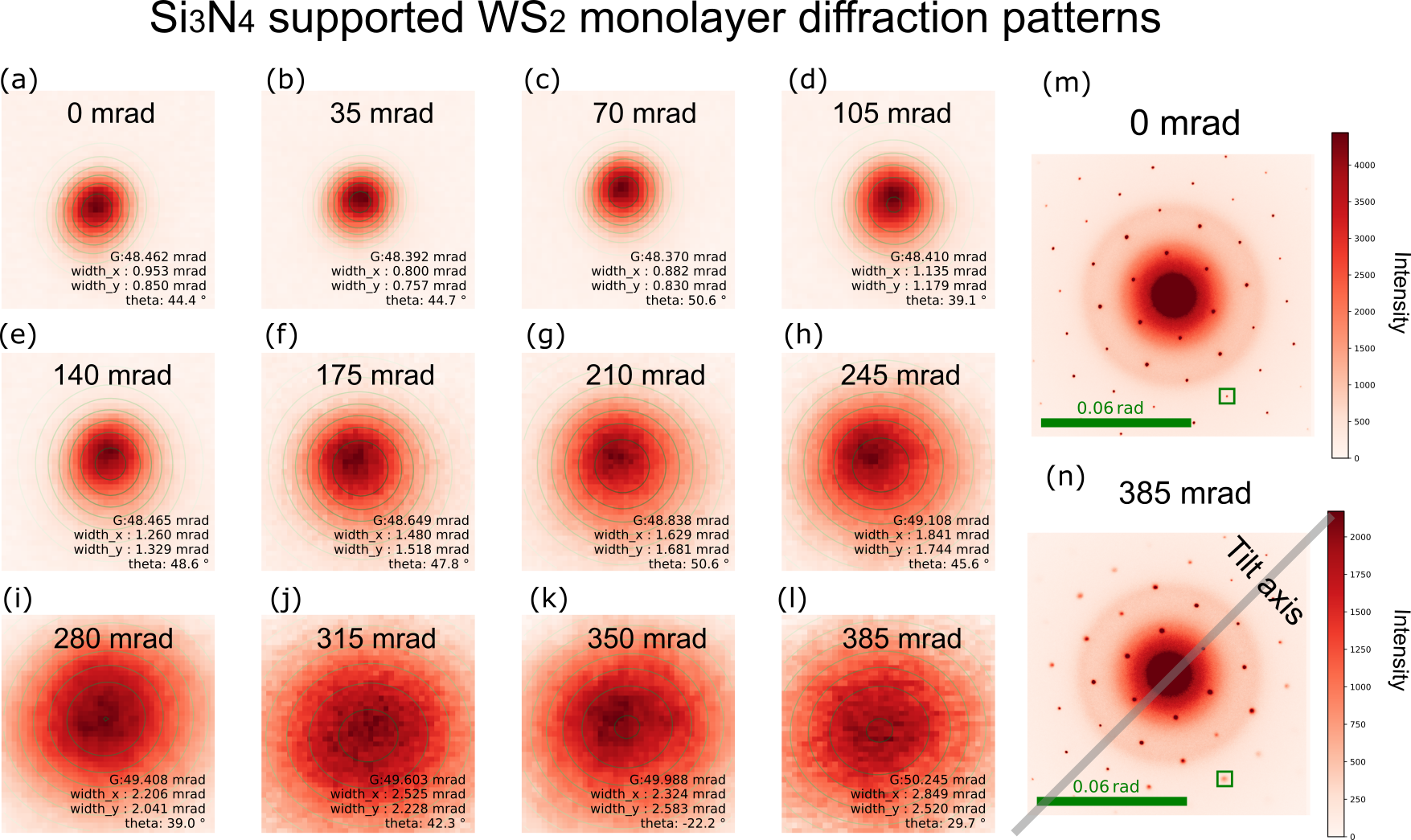}\\
  \captionsetup{font={footnotesize,stretch=1.1},  justification=justified}
  \caption{Two dimensional Gaussian fit of Si$_3$N$_4$ supported WS$_2$ monolayer diffraction pattern tilt series. (a)-(l) Two dimensional Gaussian fit of the selected peak from 0 mrad to 385~mrad sample tilt angle. The FWHM, rotation angle, and intensity are given at the bottom right corner. The candidate peak selected, $\langle\overline{1}3\overline{2}0\rangle$, is marked by a green rectangle in (m) and (n).}
      \label{Fig.SI6_Si3N4}
\end{center}
\end{figure}

\begin{figure}[H]
\begin{center}
  \includegraphics[width=0.7\linewidth]{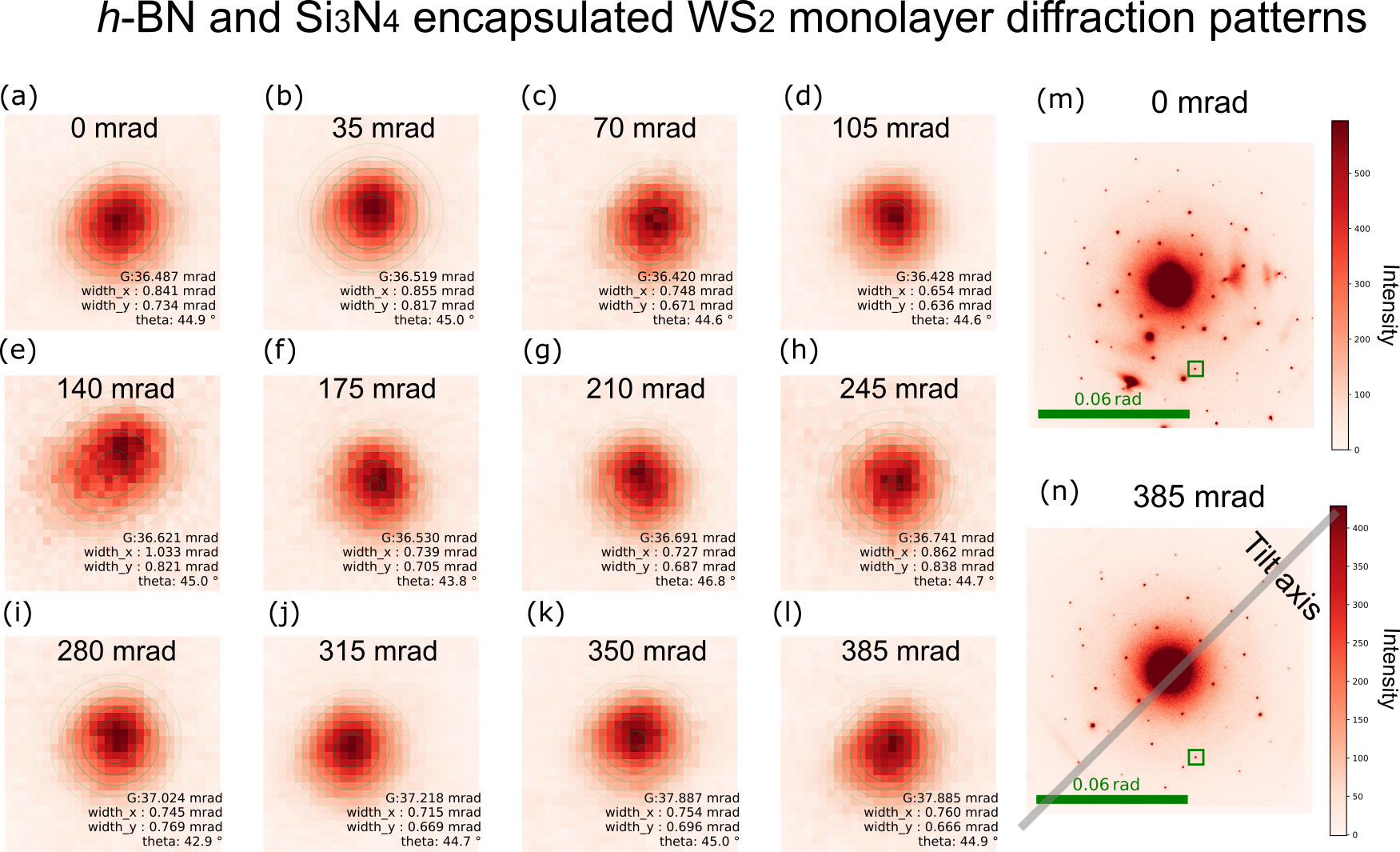}\\
  \captionsetup{font={footnotesize,stretch=1.1},  justification=justified}
  \caption{Two dimensional Gaussian fit of \textit{h}-BN and Si$_3$N$_4$ encapsulated WS$_2$ monolayer diffraction pattern tilt series. (a)-(l) Two dimensional Gaussian fit of the selected peak from 0 mrad to 385~mrad sample tilt angle. The FWHM, rotation angle, and intensity are given at the bottom right corner. The candidate peak selected, $\langle12\overline{3}0\rangle$, is marked by a green rectangle in (m) and (n).}
  \label{Fig.SI6_Si3N4_hBN}
\end{center}
\end{figure}

In Fig. \ref{Fig.4}, we plot the peak width evolution with the tilt angle, so we show which peaks we chose and how we fit the data in this section. The chosen peaks are fitted by two dimensional Gaussian fit. In Fig. \ref{Fig.SI6_ex}--\ref{Fig.SI6_Si3N4_hBN}, two dimensional Gaussian fit of the diffraction peaks in exfoliated, CVD-grown, \textit{h}-BN supported, \textit{h}-BN encapsulated, Si$_3$N$_4$ supported and \textit{h}-BN/Si$_3$N$_4$ encapsulated WS$_2$ monolayer are shown. In each figure, intensity (G), FWHM in x direction, FWHM in y direction, and rotation angle (\(\theta\)) in the bottom-right corner in (a)-(l) are denoted. The whole diffraction pattern and candidate peaks selected for fitting in (a)-(l) are plot in (m) and (n). Scale bar is only for (m) and (n) in Fig. \ref{Fig.SI6_ex}-\ref{Fig.SI6_Si3N4_hBN}, chosen region plot (a)-(l) is refined. The tilt axis is along the diagonal direction shown in (n) by the grey line. The fit of the two dimensional Gaussian is carried out with the intensity at linear scale as the reference \cite{Meyer2007}. Several roughness measurement at several different region on the same sample are made as shown in Fig. \ref{Fig.4}.

\end{document}